\documentclass[12pt,a4paper]{article}
\usepackage{epsfig}

\def\eq#1{{Eq.~(\ref{#1})}}
\def\fig#1{{Fig.~(\ref{#1})}}
\newcommand{\Le}{\left(}
\newcommand{\Ra}{\right)}

\newcommand{\beq}{\begin{equation}}
\newcommand{\eeq}{\end{equation}}   
\newcommand{\beqar}[1]{\begin{eqnarray}\label{#1}}
\newcommand{\eeqar}{\end{eqnarray}}
\def\zpc#1#2#3{    {\it Z. Phys. }{\bf C#1}:#2 (#3)}

\begin{document}

\title {{~}\\
{\Large \bf Proton-proton interaction in constituent quarks model
at LHC energies }\\}

\author{ {~}\\
{~}\\
{\large\bf S. ~B o n d a r e n k o\,\,${}^{a)}$\,
\thanks{Email: sergb@mail.desy.de},
 \quad  E.~L e v i n\,${}^{b)}$\,\thanks{E-mail: leving@post.tau.ac.il}}
 \\[10mm]
 {\it\normalsize ${}^{a)}$ II Theory Institute,}\\
 {\it\normalsize Luruper Chausse 149,}\\ 
 {\it\normalsize Hamburg University, Hamburg, Germany}\\
 [10mm]
 {\it\normalsize ${}^{b)}$ HEP Department,  School of Physics and Astronomy,}\\
 {\it\normalsize Raymond and Beverly Sackler Faculty of Exact Science,}\\
 {\it\normalsize Tel-Aviv University, Ramat Aviv, 69978, Israel}\\}
\maketitle
\thispagestyle{empty}

\begin{abstract} 
In this paper  we consider the soft processes at LHC energies in the 
framework of the Constituent 
Quark Model (CQM). We show that this rather naive model is able 
to describe  all available soft data 
at lower energies and to predict the behavior of 
the total cross section, elastic and diffractive  
cross sections at the LHC energy. It turns out that 
the ''input''   Pomeron, which has been used in this 
approach, has  parameters that  are close to so 
called ''hard''  Pomeron with rather large intercept 
$\Delta \approx 0.12$ and small value of the 
slope $\alpha'_P\,\approx\,0.08 GeV^{-2}$. 
We show that the elastic amplitude has a 
minimum at impact parameter $b =0$ and a maximum at $b 
\approx 2 GeV^{-1}$. Such a behavior is a result 
of overlapping the parton clouds that belong to 
different quarks in the hadron.

\end{abstract} 
\thispagestyle{empty}
\newpage

\section{Introduction}

   One of the most challenging problems of QCD is to find  the  correct degrees of
freedom for high energy ''soft'' interactions.  The question is
what  set of quantum numbers diagonalizes the interaction matrix at high
energies. The Constituent Quark Model (CQM) \cite{AQM} is one  of the
models which can be a good candidate for a  correct descriptions of 
the ''soft'' interactions, see also \cite{DOF}.
In this model  the constituent quarks  play roles of 
the correct degrees of freedom
for high energy QCD and  the structure of a hadron is characterized by 
two radii: the proper size of the constituent quark ($R_Q$) 
and the typical distance between two constituent quarks in a hadron ($R$).
The main assumption is that $R\,\gg\,R_Q$.

\begin{figure}[hptb]
\begin{center}
\psfig{file=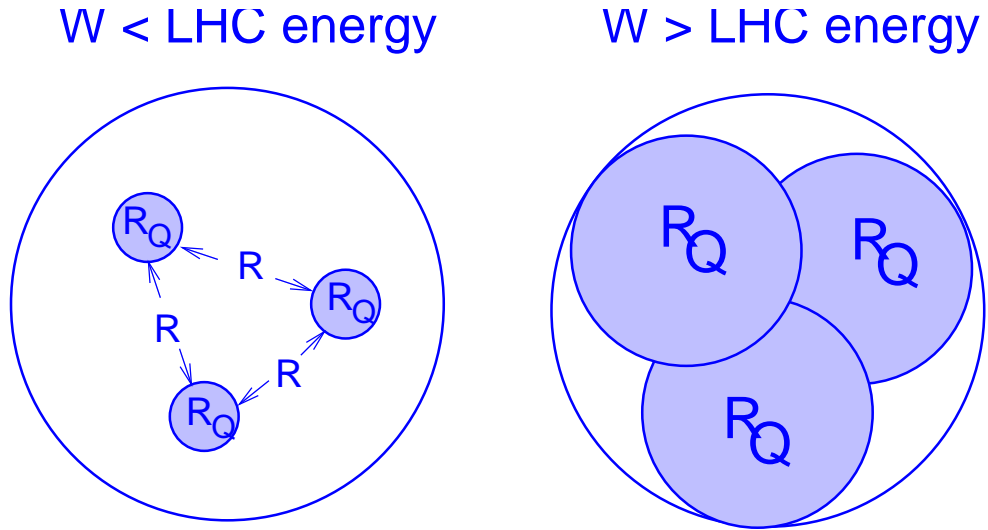,width=130mm}
\end{center} 
\caption{\it\large The proton structure at different energies in CQM.}
\label{aqmw}
\end{figure}

In spite of the fact that this model looks rather naive, it is 
supported by the 
two sets of the experimental data, namely,  CDF  double parton cross section 
at the  Tevatron, \cite{CDFDP}, and HERA data on inclusive diffraction 
production with nucleon excitation, \cite{DIFFREV,HERAREV}.
In our paper , \cite{BLN}, we examined these data and found,
that the CQM   model describes a lot of ''soft'' data in the first
approximation, see also \cite{QUARKBOOK}. The radius of the constituent quark,
which was found in Ref.\cite{BLN}, turned out to be small:
$R^2_{quark}\,\approx\,0.1\,-\,0.2\,GeV^{-2}$. However, this radius depends  
on energy
(at least logarithmically as 
$R²\,= \,R^2_{quark}\,+\,4\,\alpha_P\,\ln (W²/W²_0) $),
and the possible scenario is
that at the LHC energy this radius becomes compatible with the distance 
between constituent quarks ($R$) (see \fig{aqmw}).
Therefore we could expect a new physics at the LHC in such an approach.

  In this paper we are going to develop a systematic approach to high 
energy scattering in CQM. To our surprise we found that
only the simplest diagrams of this model has been discussed in details,
namely the diagrams which   
include only interaction between
a pair of quarks. In our model
we include all possible quark interactions, considering an 
eikonal approximation for
the scattering amplitude of two colliding quarks, (see \fig{AmP1}).
In this case, the first  contribution to the elastic amplitude  will be simply
nine interactions between constituent quarks in the  protons. 
Further contributions to the amplitude will include 
the other interaction between quarks 
(see \fig{AmP3}).
Taking into account all possible configurations
of quarks  we
will obtain amplitude which  contains nine different contributions 
with the alternating signs. Such a structure of the 
answer is similar to the  scattering amplitude of light nuclei 
(tritium-tritium scattering). 
We will show, that the result  satisfies the unitarity constraint, 
when we consider interactions between 
different configurations of the constituent quarks. 
The effects of interaction of several pairs of quarks
is especially important 
at high energies. If we will consider our amplitude at 
asymptotically very high energy, where we will may
replace eikonalized amplitude by a  step function,
the different parts of the amplitude will cancel each other and
only the last diagram, in which all quarks of the projectile interact 
with all quarks of the target,  will survive.
This last  diagram will give unitarized Froussart-like answer.
In our estimates  it turns out, that already at energies 
$\,\sqrt{s}\,=1855\,\,GeV\,\,$
we need to consider the  interaction of the five pairs of quarks
in the protons. 
At the LHC  energies, the interaction of the seven quark pairs 
is essential. This structure of the interaction
changes not only the high energy dependence 
of the scattering 
amplitude but also leads  
to a quite different  impact
parameter dependence  of the answer. 
In the case of these multi-Pomeron exchanges
between the different pairs of quarks, the amplitude has a minimum at low $b$.
Such change is very important since it could affect the behavior 
of the slope both as function of energy and as function of
momentum transfer.

   Another interesting problem, addressed in this paper, 
is the values of  parameters of the ''initial''
Pomeron. Indeed, it is widely believed, that    
the ''soft'' Pomeron is originated by the 
non-perturbative QCD contributions,  which are out of theoretical
control at the moment. Everything that we know about `soft" 
Pomeron  is a mixture of our phenomenological knowledge with the 
general theoretical ideas on the properties of non-perturbative 
QCD contributions (see Refs. 
\cite{BLT,CTM,KL,KKL,KASI,KA,CKT,CSTT,HEIDELBERG,SHURYAK,SSP,KLT}).
The question is the following,
can we obtain the well known phenomenological
Pomeron using as ''initial'' the  Pomeron with the large intercept and small slope,
i.e  so called ''hard'' Pomeron, \cite{FL} ?
Does the ''hard'' Pomeron play any role in the ''soft'' interactions?  There 
exist
two different points of view on this question. The first one is that 
we need to introduce
separately two objects, ''soft'' and ''hard'' Pomerons,  which have different 
properties and
contribute differently, each in a different  kinematic region, see for example
\cite{BLT,CTM,KASI,CSTT} and references therein.
Another point of view, see for example \cite{MOT}, is that non-perturbative 
physics comes in our calculations
only in the form of the  boundary and/or initial  conditions and 
the ''soft'' Pomeron arises 
as a result 
of unitarization effects and self-interactions of the ''hard'' Pomerons 
in the amplitude. Our model may help to clarify the situation.  
Taking into account all effects of the unitarization,
eikonalization and accounting all 
interactions between the different configurations of quarks,
we will fit the experimental data.  The fit will determine
the parameters of the ''initial'' Pomeron , such as its intercept and slope, 
as well as the radius of the constituent quark.
We show in this paper, that the  parameters 
of the initial ''soft'' Pomeron are close to the  parameters of
the ''hard'' Pomeron and quite different from the parameters 
of the Donnachie-Landshoff Pomeron \cite{DL}, which 
is usually considered as a typical ''soft'' Pomeron.

The structure of the paper is as follows.
In Section 2  
we discuss in more details our approach and methods of the 
calculations. 
In Section 3 we apply our model to the 
p-p data  and fit the experimental data in order to find 
numerical values of the parameters of the  ''initial'' Pomeron.
Section 4 is dedicated to the elastic amplitude as a function of the
impact parameter.
In this section we also consider the different contributions to the elastic amplitude  
due to interactions of different numbers of quark-quark pairs.
In  Section 5 we calculate the survival probability (SP) of the 
exclusive hard processes in p-p scattering . 
The cross section of the diffractive dissociation process is calculated 
in the Section 6.
The last section, Conclusion, contains the  main results of the paper as 
well as the discussion of  
a  future work in the proposed direction.

\section{ Proton-proton scattering in the Pomeron approach }

\begin{figure}[hptb]
\begin{center}
\psfig{file=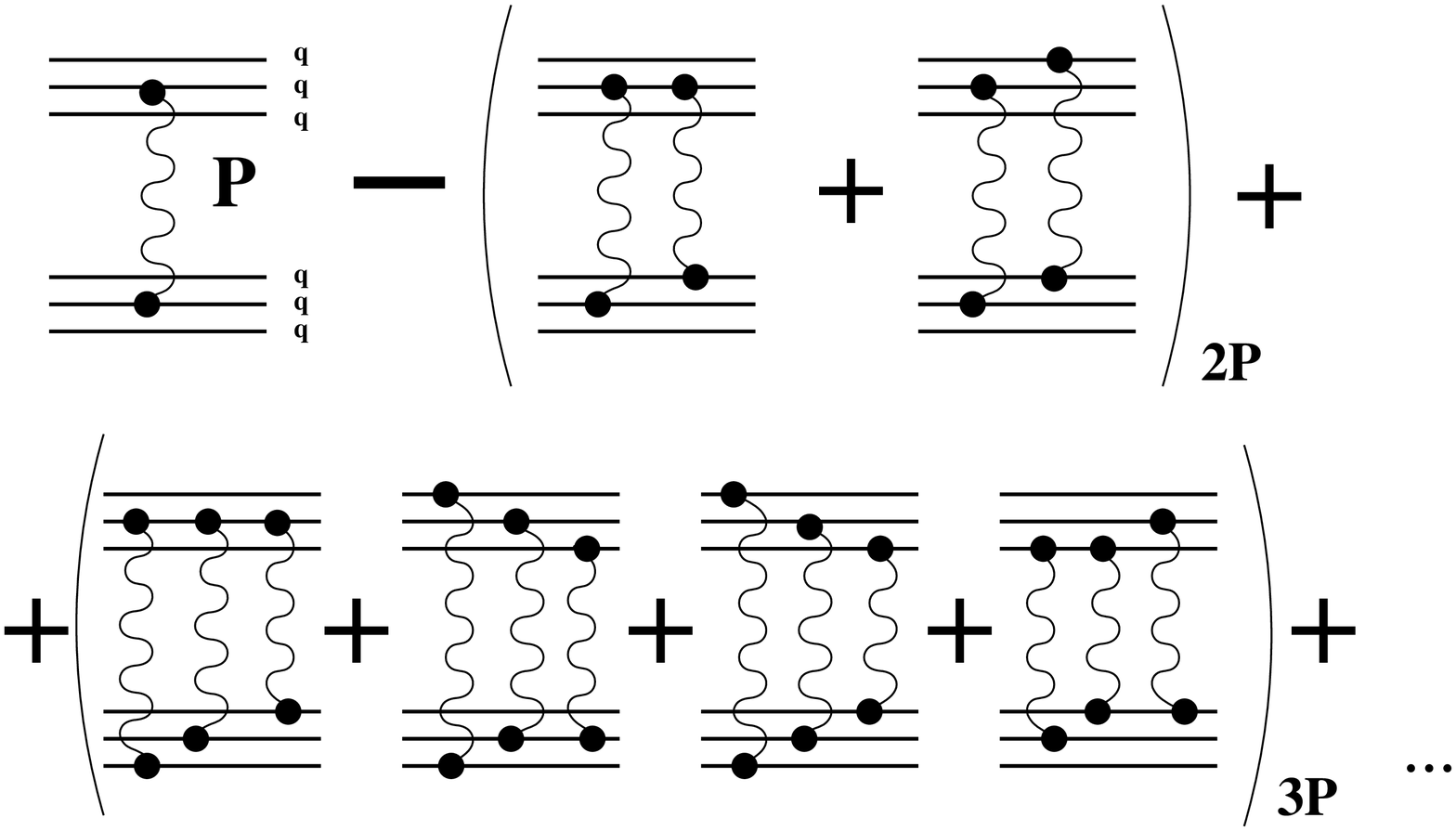,width=140mm}
\end{center} 
\caption{\it\large The three first orders of the p-p elastic amplitude
in the CQM.}
\label{AmP3}
\end{figure}

The key ingredient of the CQM is the quark-quark scattering
amplitude. Considering this model, 
we need to determine  the form
of the single Pomeron exchange between two quarks. The next step will be the 
eikonalization of the single scattering amplitude, that means a replacement 
of the single Pomeron  exchange in the  scattering of the  
particular pair of 
quarks by the eikonal
amplitude. The third step in our calculation will be
the consideration of  the interactions between  
 all possible quark configurations
in colliding protons, see Fig.~\ref{AmP3}.
For this last step we
need to know the wave function of the 
quarks in the proton and the
vertices of the Pomerons-quarks interactions. Only after 
the determination of the    
wave function and vertices we will be able
to calculate the diagrams for the quark-quark interactions.

 Now let us consider
these  problems step by step. 
 
\subsection{Quark-quark interactions}

\begin{figure}[hptb]
\begin{center}
\psfig{file=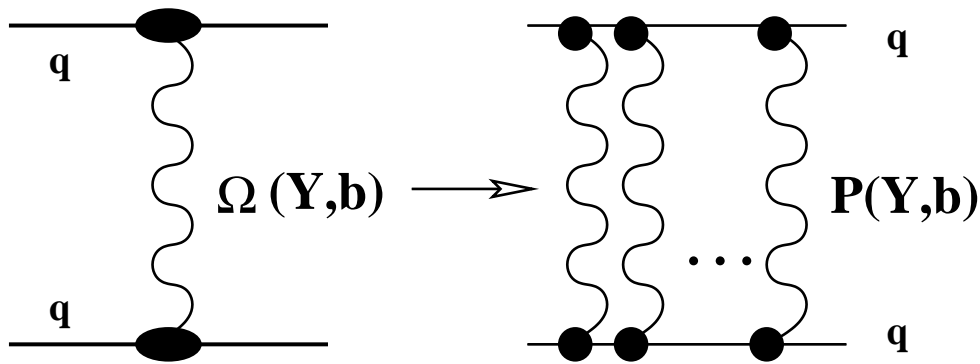,width=130mm}
\end{center} 
\caption{\it\large The Pomeron exchange  and eikonalized 
quark-quark amplitudes.}
\label{AmP1}
\end{figure}

 We determine  the amplitude for q-q and p-p scattering in the impact parameter 
representation. In this case the ''soft'' Pomeron  exchange 
for the interaction of the pair of the quarks (Pomeron propagator)
has the following form (see more about soft Pomeron in 
Ref.\cite{KA,DL,POM}:

\beq\label{SinPom}
\Omega_{q-q}(Y,b)\,=\,\sigma_{0}\,e^{\Delta\,Y}\,\frac{e^{-b^2/R^2}}{\pi\,R^2}\,,
\eeq
here, $\,Y=\ln\,(s/1\,GeV^{2})\,$ is the rapidity of the process,
b is the impact parameter of the process, $\,\Delta\,$ is the intercept
of the 'initial', input Pomeron and 
\beq\label{RadQ}
R^2=8\,R^{2}_{Q}\,+\,4\,\alpha^{'}_P\,\ln\,(s/s_0)\,\,. 
\eeq
Here $\,R^{2}_{Q}\,$ is the squared radius of the constituent quark
and $\,\alpha^{'}_P\,$ is the slope of the  input Pomeron trajectory.
The numerical values of the $\,\sigma_{0}\,$,  
$\,\Delta\,$, $\,R^{2}_{q}\,$
and $\,\alpha^{'}_P\,$ we will find   
fitting data for p-p scattering.
The eikonal amplitude, which is
a ''main'' ingredient in  our calculations, we determine
as follows:

\beq\label{EikPom}
P_{q-q}(Y,b)\,=\,1\,-\,e^{-\Omega_{q-q}(Y,b)\,/\,2}\,\,,
\eeq
see Fig.~\ref{AmP1}, where $P_{q-q}$ is the imaginary part of the 
quark-quark scattering amplitude at high energy.
Below , discussing  the single quark-quark interaction
amplitude,
we will mean only the amplitude given by  \eq{EikPom}.
The Pomeron of \eq{SinPom} we will consider only as 
the ''input'', initial Pomeron of our problem.

\subsection{The model of the proton}

\begin{figure}[hptb]
\begin{center}
\psfig{file=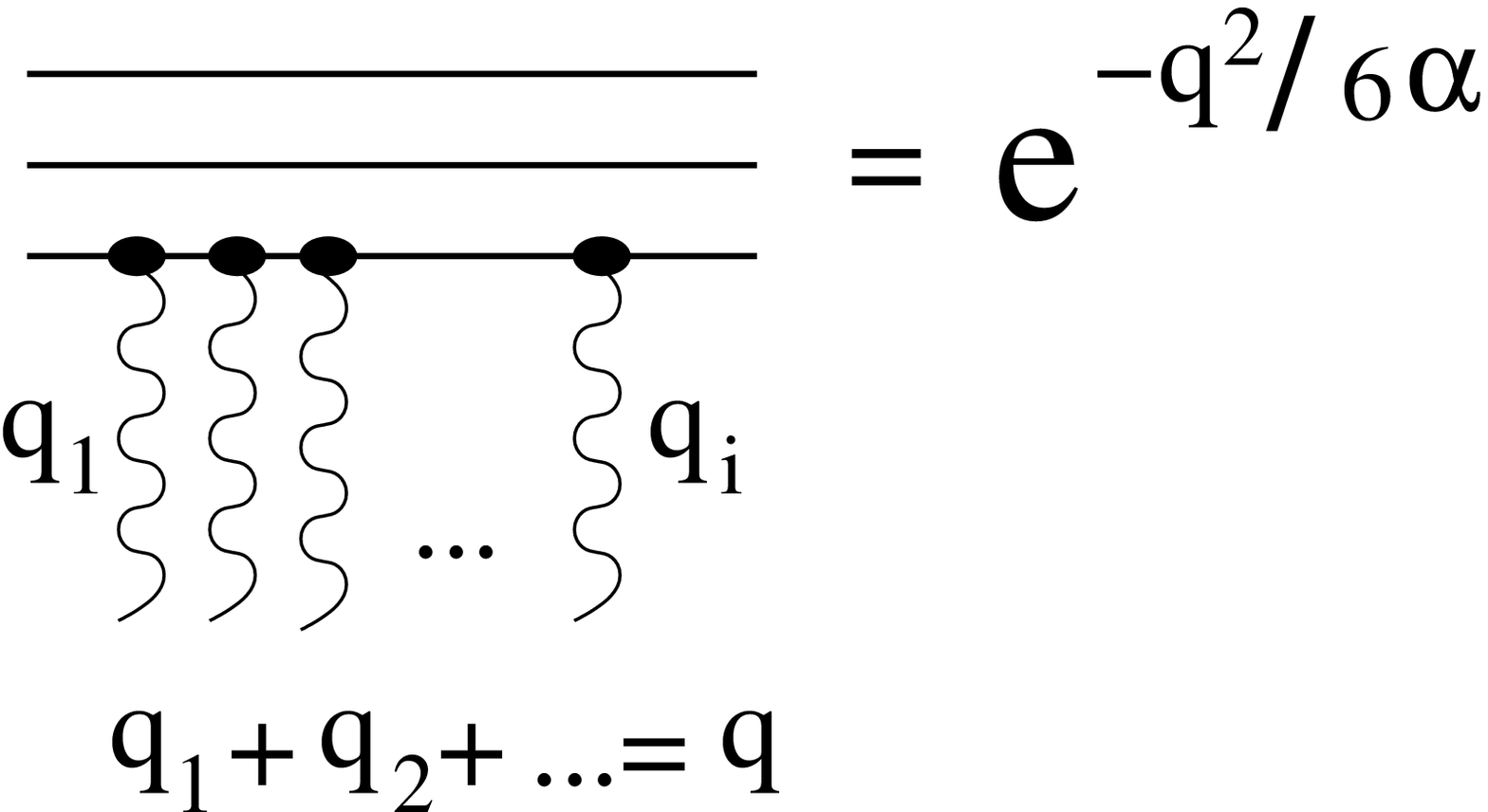,width=100mm}
\end{center} 
\caption{\it\large The one Pomeron vertex in quark-quark interaction.}
\label{AmP4}
\end{figure}

\begin{figure}[hptb]
\begin{center}
\psfig{file=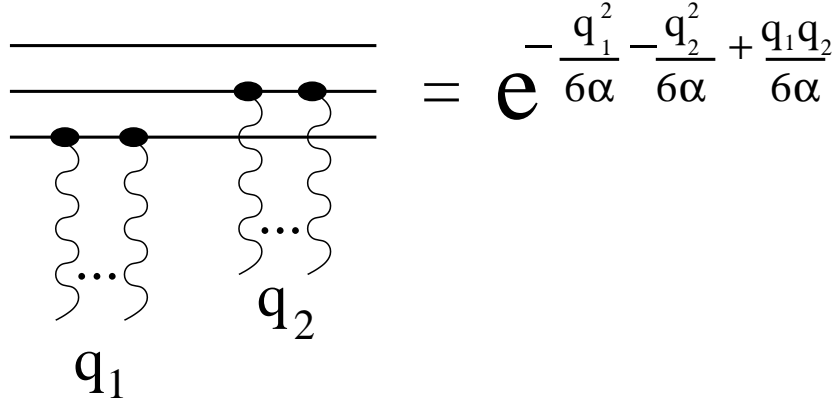,width=110mm}
\end{center} 
\caption{\it\large Two Pomeron vertex in quark-quark interaction.}
\label{AmP5}
\end{figure}

\begin{figure}[hptb]
\begin{center}
\psfig{file=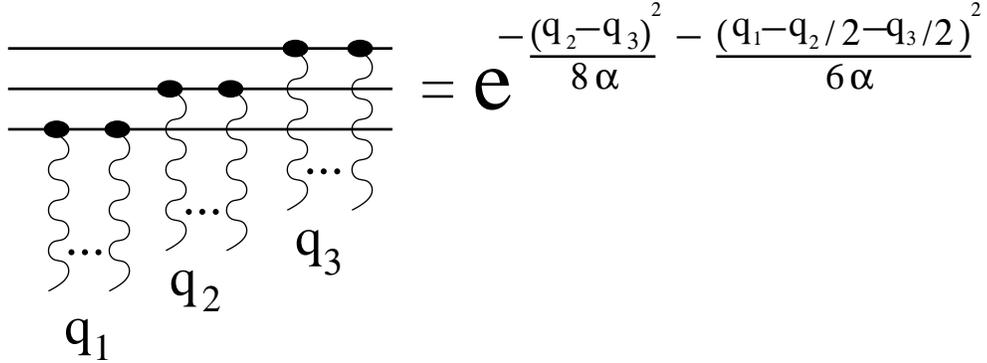,width=130mm}
\end{center} 
\caption{\it\large The three Pomeron vertex in q-q interaction.}
\label{AmP6}
\end{figure}

 In order to take into account all possible configurations
of the interacting pairs  of the quarks, 
see Fig.~\ref{AmP3} and all figures in the
Appendices,
we need to know the analytical expressions for the vertices of the quark-Pomeron
interactions. It is  clear, 
that we need  to calculate only three types of such vertices, 
see Fig.~\ref{AmP4}-Fig.~\ref{AmP6}, 
where there are one, two or three groups of the 
Pomerons  attached to the one, two
or three quarks. In order to 
calculate these vertices 
we need to know  the wave functions of the constituent
quarks inside a proton. We use a very simple Gaussian model for this
wave function, which corresponds to the oscillatory potential between pair of quarks 
in a proton. In this model 
we have ( see  \cite{DFK}):

\beq
\Psi\,=\,\frac{\alpha}{\pi\,\sqrt{3}}\,
e^{-\frac{\alpha}{2}\Le\,\sum\,x^{2}_{i}\,\Ra}\,
\eeq 
where the constant $\,\alpha\,$ is related  to
the electromagnetic radius of the proton:

\beq
\alpha\,=\,1/R^{2}_{electr}\,\approx\,0.06\, GeV^{2}\,\,.
\eeq

For the diagram of  Fig.~\ref{AmP4}, we have:

\beq\label{Vert1}
V_{1}(q)\,=\,\int\,|\Psi(x_1,x_2,x_3)|^{2}\,
\delta(\vec{x}_1+\vec{x}_2+\vec{x}_3)\,
e^{i\,\vec{q}\,\vec{x}_1}\,=\,e^{-\frac{q^2}{6\,\alpha}}\,.
\eeq 

We determine  the impact factor for  two groups of 
Pomerons attached to  two different quarks (see Fig.~\ref{AmP5}) : 

\beq\label{Vert2}
V_{2}(q_1,q_2)\,=\,\int\,|\Psi(x_1,x_2,x_3)|^{2}\,
\delta(\vec{x}_1+\vec{x}_2+\vec{x}_3)\,
e^{i\,\vec{q}_1\,\vec{x}_1\,+\,\vec{q}_2\,\vec{x}_2\,}\,
=\,e^{-\frac{q_{1}^{2}}{6\,\alpha}-\frac{q_{2}^{2}}{6\,\alpha}+
\frac{\vec{q}_{1}\,\vec{q}_{2}}{6\,\alpha}}\,.
\eeq 

The last impact factor is the vertex which is shown in 
Fig.~\ref{AmP6}. We have for this vertex :
 
\beq\label{Vert3}
V_{3}(q_1,q_2,q_3)\,=\,\int\,|\Psi(x_1,x_2,x_3)|^{2}\,
\delta(\vec{x}_1+\vec{x}_2+\vec{x}_3)\,
e^{i\,\vec{q}_1\,\vec{x}_1\,+\,\vec{q}_2\,\vec{x}_2\,+
\,\vec{q}_3\,\vec{x}_3\,}\,
=\,e^{-\frac{(\vec{q}_{2}\,-\,\vec{q}_{3})^{2}}{8\,\alpha}-
\frac{(\vec{q}_{1}-\vec{q}_{2}/2-\vec{q}_{3}/2)^{2}}{6\,\alpha}}\,.
\eeq
In  all three  cases
the total transferred  momentum of the diagrams is defined
as the sum of  transverse momenta of all Pomerons ($q_i$):
$\,k\,=\,\sum\,q_{i}\,$.

\subsection{The elastic amplitude of p-p scattering }

   We have all ingredients for the  calculation of  the elastic amplitude.
In the CQM  we  can write the  amplitude as the sum
of the amplitudes with the different number of interacting quark pairs:

\beq\label{Ampli1}
A(s,b)\,=\,A_{1pair}(s,b)-A_{2pairs}(s,b)+A_{3pairs}(s,b)+\,...\,+\,A_{9pairs}(s,b).
\eeq
Here the amplitude for one quark pair interaction 
$A_{1pair} $ is  equal to $P_{q-q}$ which is defined by
\eq{EikPom}. 
The maximum number of possible quark pair  interactions in the amplitude
is 9. The three first orders of the possible
configurations of interactions are shown in  Fig.~\ref{AmP3}
and the calculations of  all other orders are presented in the Appendix B.
The amplitude defined in a  such way incorporates unitarity by construction.
Indeed, checking the expressions for the different term of
the amplitude, that  are written in Appendix B,  it is easy to see, that
at asymptotically high energies only the last term of  \eq{Ampli1}
will survive,  giving the Froussart-like answer for the whole amplitude.

 To  calculate the contribution of the first diagram of   
Fig.~\ref{AmP3}, i.e. the
first term of the  r.h.s. of \eq{Ampli1}.
we make Fourier transform
of the vertex of \eq{Vert1} from momentum to impact parameter space:

\beq\label{For1}
\hat{V}_{1}(b)\,=\,\int\frac{d^2\,q}{4\,\pi^2}\,e^{-q^2/(6\alpha)+
i\,\vec{q}\,\vec{b}}\,=\,
\frac{3\,\alpha}{2\,\pi}e^{-\frac{3\alpha}{2}\,b^2}\,.
\eeq
Using this vertex we obtain the contribution to the elastic amplitude 
from the  one Pomeron exchange $\,A_{1 q-q}(s,b)\,$, which is :

\beq\label{Ampli2}
A_{1pair}(s,b)\,=9\,\frac{9\,\alpha^2}{4\,\pi^2}
\int\,d^2\,b_{1}\,\int\,d^2\,b_{2}\,
e^{-\frac{3\alpha\,}{2}(\vec{b}-\vec{b}_{1})^2}\,
P_{q-q}(Y,b_{2})\,
e^{-\frac{3\alpha}{2}\,(\vec{b}_{1}-\vec{b}_{2})^2}\,\,,
\eeq
where the first coefficient in \eq{Ampli2} is 
the total number of q-q interactions
in  this order. Using more complicated vertices, which are given by
\eq{Vert2} and \eq{Vert3},  we calculate
all terms that contribute to the elastic amplitude, i.e.
terms on the r.h.s. of \eq{Ampli1}. In Appendices A,B 
the resulting expressions for the amplitude 
$\,A(s,b)\,$ are  written as well  as the examples  
of the calculations of the diagrams with the different
numbers  of  interacting pairs of  quarks.

\subsection{The total, elastic  cross sections
and the elastic slope $B_{el}$ of the proton-proton interaction}

 The expression for the amplitude (Appendix B),
is determined in the impact parameter space 
and now we can easily calculate  the 
different cross sections for the processes of interest.

 $ \mathbf{\sigma}_{\mathbf{tot}}$.   \,\,\, The total cross section in impact 
parameter representation is simply

\beq\label{Tot}
\sigma_{tot}(s)\,=\,2\,\int\,d^2\,b\,Im\, A(s,b)\,.
\eeq
Fitting the experimental data at low energies
we  also add to this cross section  
the contribution of 
the secondary Reggeons, $\,\sigma_{tot}^{Reg}\,$, see \cite{SPR}.

 $ \mathbf{\sigma}_{\mathbf{el}}$.   \,\,\,
The elastic cross section in the same  framework is equal:

\beq\label{El}
\sigma_{el}(s)\,=\,\int\,d^2\,b\,|A(s,b)|^2\,\,=\,\,\int\,d^2\,b\,(\,Im\,A(s,b)\,)^2.
\eeq
where we assumed that the quark-quark amplitude is mostly imaginary as it follows 
from the eikonal approximation.

 $\mathbf{B}_{\mathbf{el}}$.\,\,\, We
consider only the first term of the elastic 
amplitude, i.e $\,A_{1 q-q}(s,b)\,$. 
In this case we have for the slope $\,B_{el}\,$:

\begin{eqnarray}\label{Slope}
B_{el}\,& = &\,
\,\frac{\frac{9\,\alpha^2}{4\,\pi^2}\,
\int\,d^2\,b_{1}\,\int\,d^2\,b_{2}\,\int\,d^2\,b\,\,b^{2}\,\,
e^{-\frac{3\alpha\,}{2}(\vec{b}-\vec{b}_{1})^2}\,
P_{q-q}(Y,b_{2})\,
e^{-\frac{3\alpha}{2}\,(\vec{b}_{1}-\vec{b}_{2})^2}\,}
{2\,\int\,\,d^2\,b\,P_{q-q}(Y,b)}\,=\,\\
\,&=&\,
\frac{1}{\alpha}\,\frac{\sigma_{tot}^{1 q-q}}{\sigma_{tot}^{1 q-q}}\,+\,
\frac{\int\,\,d^2\,b\,\,b^2\,\,P_{q-q}(Y,b)}{\sigma_{tot}^{1 q-q}}\,.
\end{eqnarray}
Here 
$$
\sigma_{tot}^{1 q-q}\,=\,2\,\int\,d^2\,b\,P_{q-q}(Y,b)\,.
$$
We  add the contribution of
the secondary Reggeons to this expression , which is needed 
to be taken into account at low energies:

\beq\label{Slope1}
B_{el}\,=\,\frac{9\,\sigma_{tot}^{1 q-q}}
{9\,\sigma_{tot}^{1 q-q}+\sigma_{tot}^{Reg}}\,R_{electr}^{2}\,+\,
\frac{9\,\int\,\,d^2\,b\,\,b^2\,\,P_{q-q}(Y,b)}
{9\,\sigma_{tot}^{1 q-q}+\sigma_{tot}^{Reg}}\,+\,
2\,(\alpha^{'}_{R}-\alpha^{'}_{SP})\,\ln\,(s/s_0)\,
\frac{\sigma_{tot}^{Reg}}{9\,\sigma_{tot}^{1 q-q}+\sigma_{tot}^{Reg}}\,.
\eeq
The third term of the r.h.s of \eq{Slope1} contains
the contribution of the secondary Reggeons, which we do not consider
in our model. The secondary Reggeons we account only on the
level of proton-proton scattering.
Therefore, the parameters which we take for
this contribution have a pure phenomenological origin. 
We take for the slope of secondary Reggeons
$\,\alpha^{'}_{R}\,=\,0.86\,GeV^{-2}\,$ and 
for the slope of phenomenological ''soft'' Pomeron
\footnote{The slope $\,\alpha^{'}_{SP}\,$ is the slope
of the phenomenological Pomeron in proton-proton scattering , 
see \cite{DL,POM,SPR},
and it is not related to the slope  $\,\alpha^{'}_{P}\,$
of our ''initial'' Pomeron in the quark-quark scattering.} 
$\,\alpha^{'}_{SP}\,=\,0.25\,GeV^{-2}\,$ at 
$\,\sqrt{s}_0\,=\,9\,GeV\,$ in the  r.h.s of \eq{Slope1},
see \cite{DL,POM,SPR}.
Generalizing this expression to the case of
full  elastic  amplitude ( see \eq{Ampli1}) we obtain 
at  low energies: 

\begin{eqnarray}\label{Slope2}
B_{el}\,&=&\,\frac{9\,\sigma_{tot}^{1 q-q}}{\sigma_{tot}+
\sigma_{tot}^{Reg}}\,R_{electr}^{2}\,+\,
\frac{9\,\int\,\,d^2\,b\,\,b^2\,\,P_{q-q}(Y,b)}
{\sigma_{tot}+\sigma_{tot}^{Reg}}\,+\,\\
\,&+&\,
2\,(\alpha^{'}_{R}-\alpha^{'}_{SP})\,\ln\,(s/s_0)\,
\frac{\sigma_{tot}^{Reg}}{\sigma_{tot}+\sigma_{tot}^{Reg}}\,+\,
\sum_{i=2}^{9}\frac{\int\,d^2\,b\,\,b^2\,\,Im\,A_{i}(b,s)}
{\sigma_{tot}+\sigma_{tot}^{Reg}}\,,
\end{eqnarray}
whereas at high energy we have

\beq\label{Slope3}
B_{el}\,=\,\frac{9\,\sigma_{tot}^{1 q-q}}{\sigma_{tot}}\,
R_{electr}^{2}\,+\,\frac{9\,\int\,\,d^2\,b\,\,b^2\,\,P_{q-q}(Y,b)}
{\sigma_{tot}}\,+\,
\sum_{i=2}^{9}\frac{\int\,d^2\,b\,\,b^2\,\,Im\,A_{i}(b,s)}
{\sigma_{tot}}\,.
\eeq

  It is also interesting to calculate the elastic cross section
using simple expression for the elastic cross section which 
is obtained in the model
with the phenomenological ''soft'' Pomeron.
Indeed, in this case we calculate the elastic cross section
using the following formula:

\beq\label{El1}
\sigma_{el}(s)\,=\,\frac{\sigma_{tot}^2}{16\,\pi\,B_{el}}\,.
\eeq 
This expression is correct only in the case when one ''soft''
Pomeron is considered. It will be interesting
to compare the calculations 
of elastic cross section given by 
\eq{El}
with  the calculations of \eq{El1}.
Indeed, in this case we check the possibility   
to reproduce the simple result of \eq{El1}
by the theory where  
many eikonalized Pomeron  exchanges are taken
into account. So, in the next section we
will perform the data fitting and will 
make the  calculations using  both expressions, 
\eq{El} and \eq{El1} 
in order to show the importance of many quark pairs interaction.

\section{The proton-proton scattering data and the 
parameters of the input Pomeron}

\begin{figure}[hptb]
\begin{tabular}{ c c}
\psfig{file=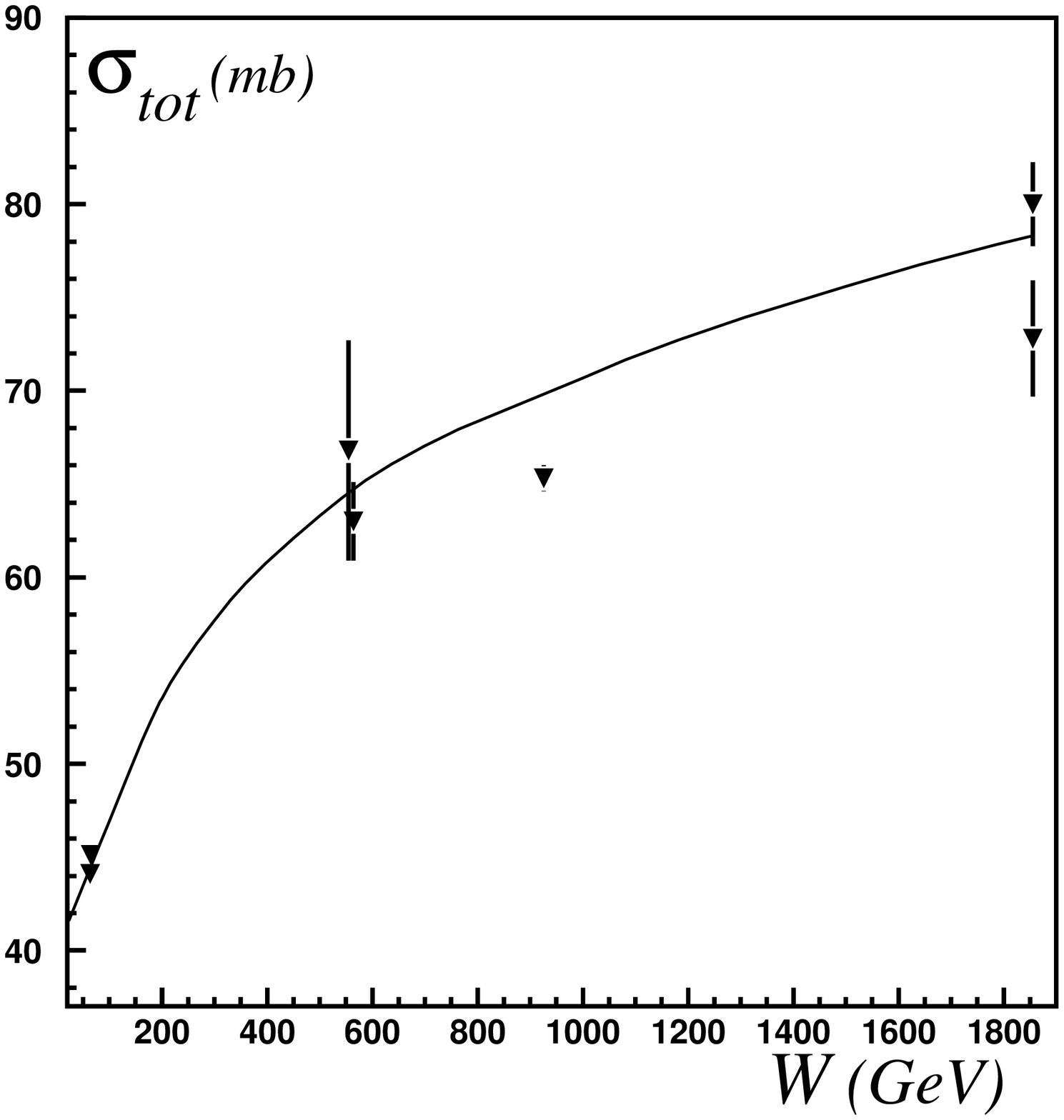,width=80mm} & 
\psfig{file=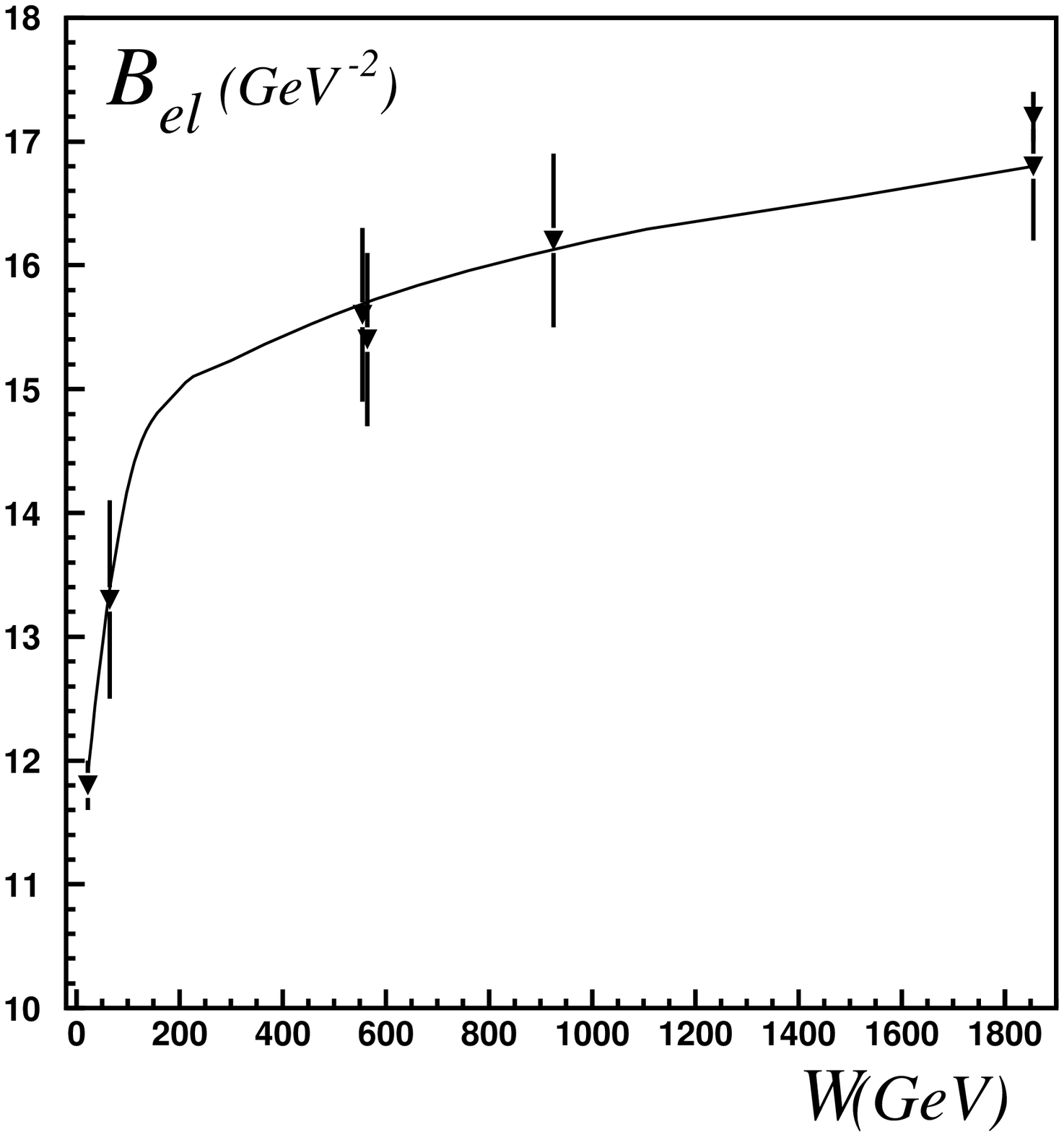,width=80mm}\\
       &  \\ 
\fig{Pr3}-a &\fig{Pr3}-b \\
 &  \\
\psfig{file=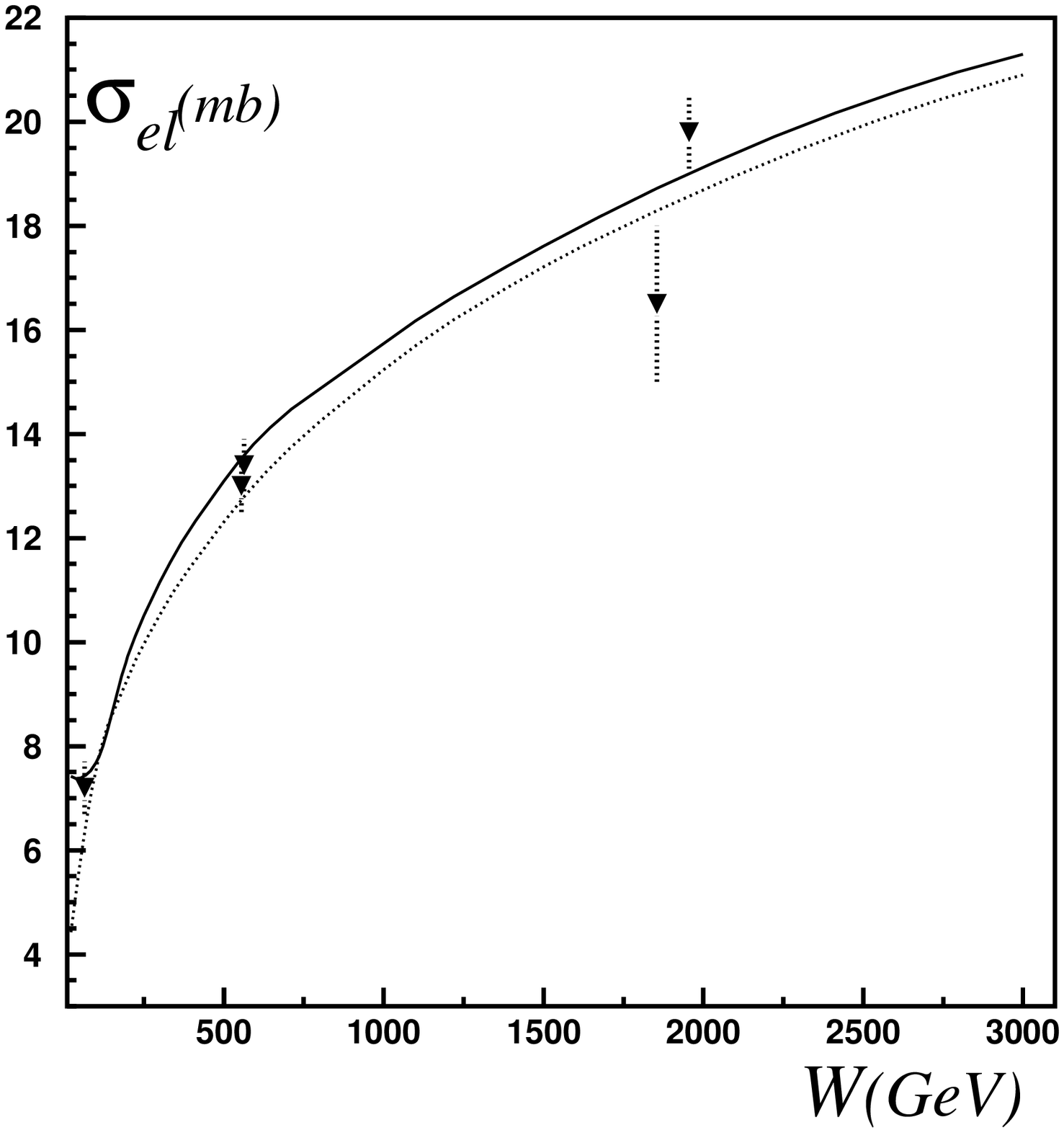,width=80mm}\\
       &  \\ 
\fig{Pr3}-c \\
 &  \\

\end{tabular}
\caption{\it The plots for the 
total cross section, \fig{Pr3}-a, elastic cross section , \fig{Pr3}-c,
and elastic slope, \fig{Pr3}-b. The experimental data
are from \cite{SPR} and  references therein. }
\label{Pr3}
\end{figure}

  Now, we are able to  apply our  model to the p-p interactions and, fitting  
the experimental observables,  we will
extract the 
values of the parameters of our input Pomeron ( see
\eq{SinPom}). In  Fig.~\ref{Pr3}
we present the plots for the total cross section, elastic cross section
and elastic slope in  $\,W=23-1855\,\,GeV\,$ energy range. 
We perform all calculations numerically
\footnote{The  Fortran code can be sent after the request 
(email:sergb@mail.desy.de)} using the  formulae of the previous subsection  
with the amplitude  written in  the  Appendix B.
There are two different plots
which we present for elastic cross section using 
definitions \eq{El} and \eq{El1}. The solid line 
represents the elastic cross section
given by \eq{El1} and dashed line represents calculations
performed with \eq{El}.

\begin{figure}[hptb]
\begin{tabular}{ c c}
\psfig{file=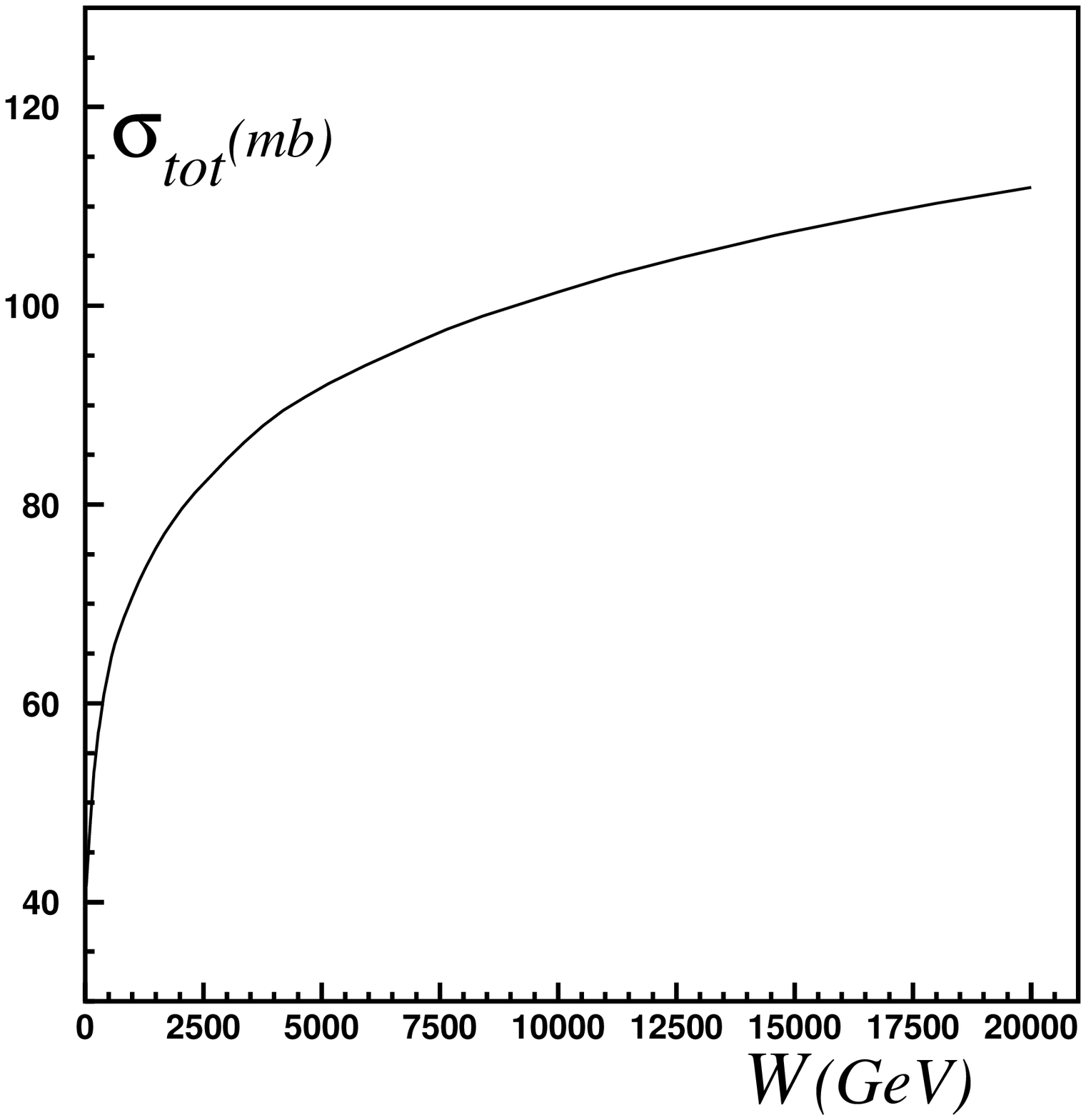,width=80mm} & 
\psfig{file=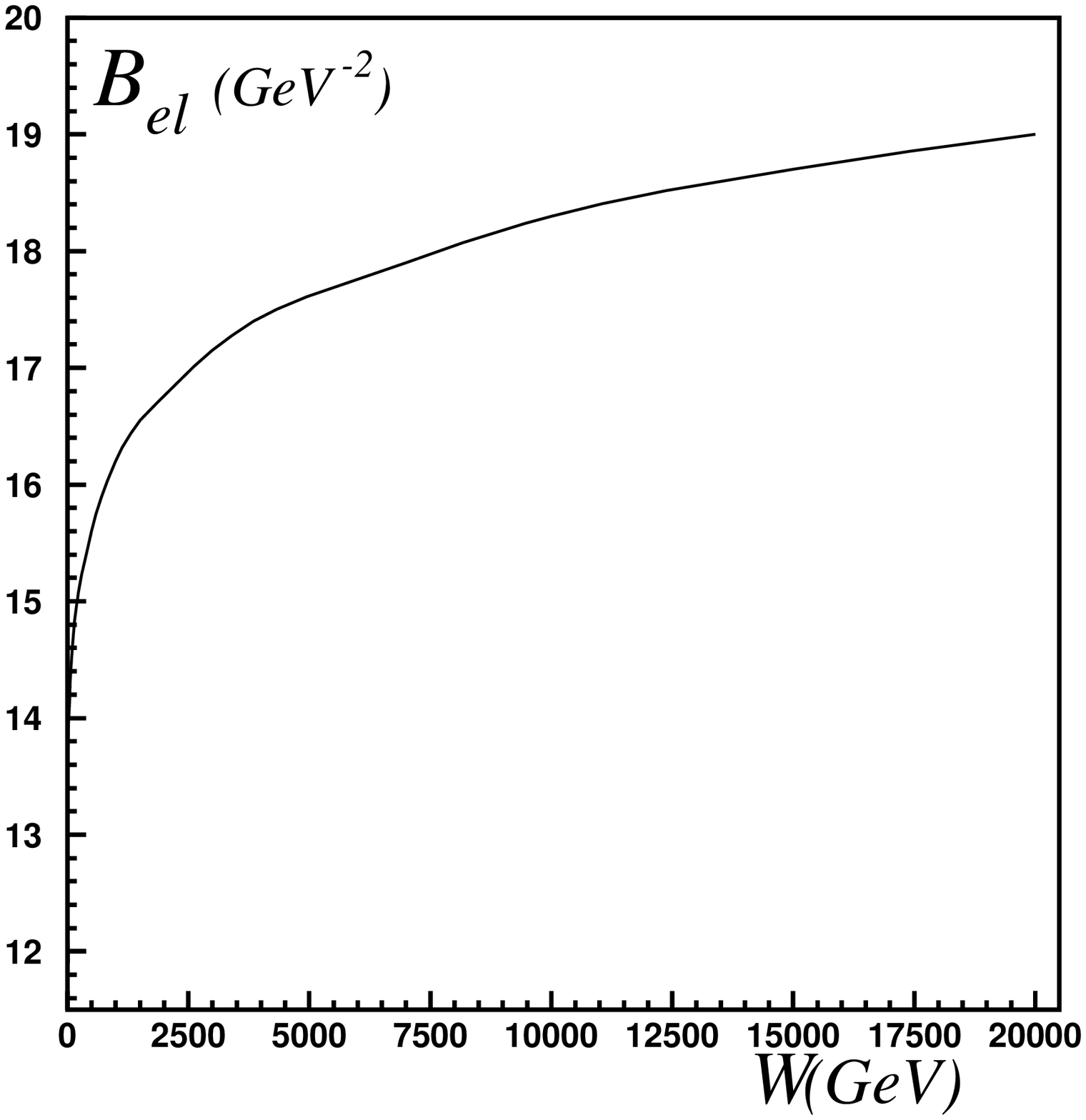,width=80mm}\\
       &  \\ 
\fig{Pr4}-a & \fig{Pr4}-b \\
 &  \\
\psfig{file=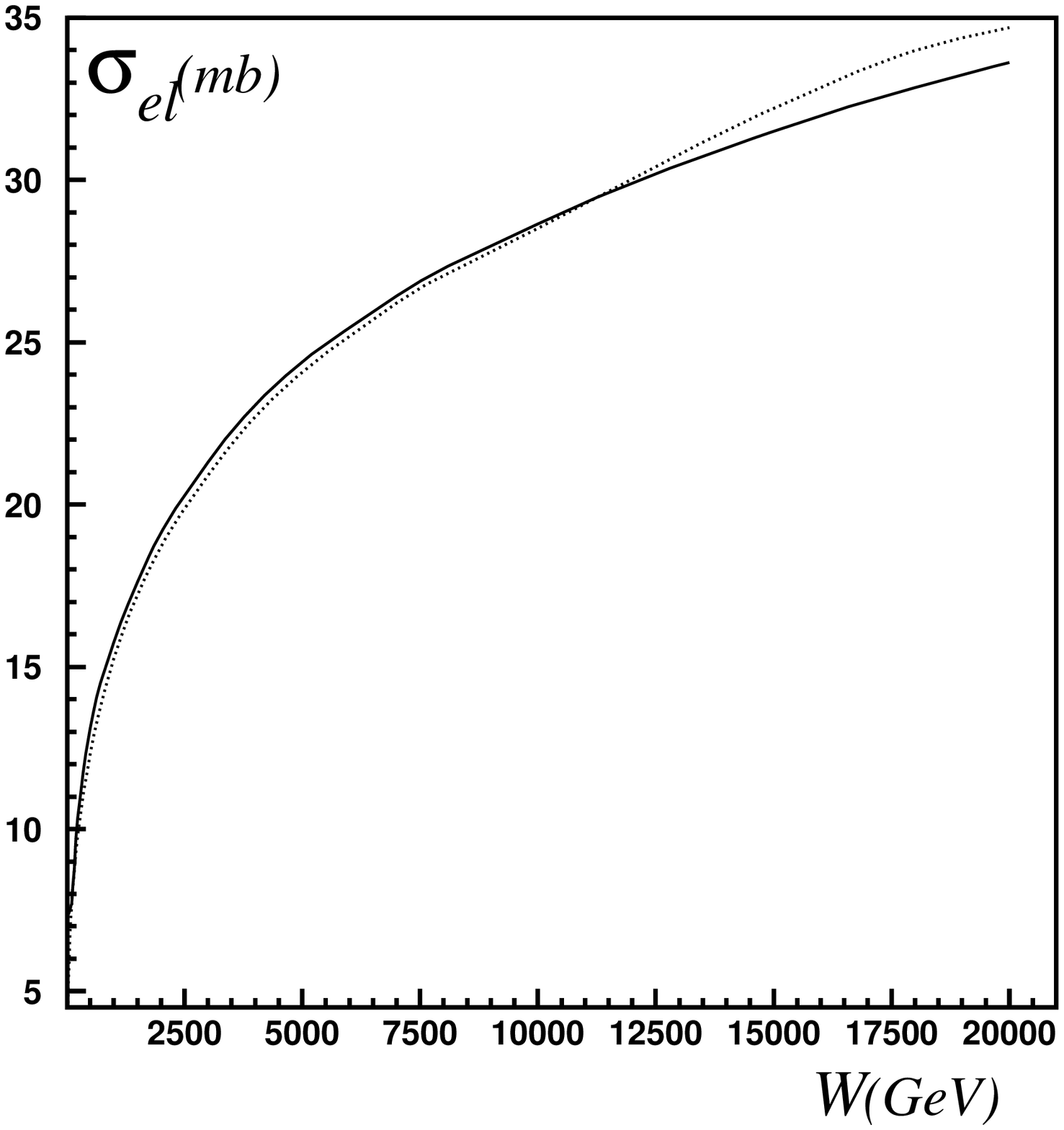,width=80mm}\\
       &  \\ 
\fig{Pr4}-c \\
 &  \\

\end{tabular}
\caption{\it The plots for the 
total cross section, \fig{Pr4}-a, elastic cross section , \fig{Pr4}-c,
and elastic slope, \fig{Pr4}-b, at high energies.}
\label{Pr4}
\end{figure}

From these  plots we see, 
that the model describes the experimental data quite well.
The parameters of the single Pomeron of 
\eq{SinPom} extracted from the data fitting
are the following:

\begin{itemize}

\item the slope of the input  Pomeron trajectory:
\beq
\alpha^{'}\,=\,0.08\,GeV^{-2};
\eeq

\item   input Pomeron's intercept:
\beq
\Delta\,=\,0.118\,;
\eeq 

\item the  
value of cross section  for  the input Pomeron at $s_0=1\,\,GeV^{2}$ :
\beq
\sigma_0\,=\,6.3\,GeV^{-2}\,; 
\eeq

\item the radius of the constituent quark:
\beq
R^{2}_{quark}\,=\,0.16\,\,GeV^{-2}. 
\eeq

\end{itemize}
With these  parameters  
we extrapolate our  calculations for the  cross sections and slope
at higher energies. The resulting plots  
are shown  in  Fig.~\ref{Pr4}.

 It should be stressed  that the above parameters are quite different from 
the Donnachie-Landshoff Pomeron \cite{DL}. The Pomeron
intercept, which we obtained, is higher and the slope is much lower than 
the intercept and slope of the  
D-L Pomeron. We may conclude, that such  small value of the Pomeron slope
indicates that our input  Pomeron can have a ''hard'' origin.

\section{Behavior of the elastic amplitude of p-p scattering
in our model}

\begin{figure}[hptb]
\begin{tabular}{ c c}
\psfig{file=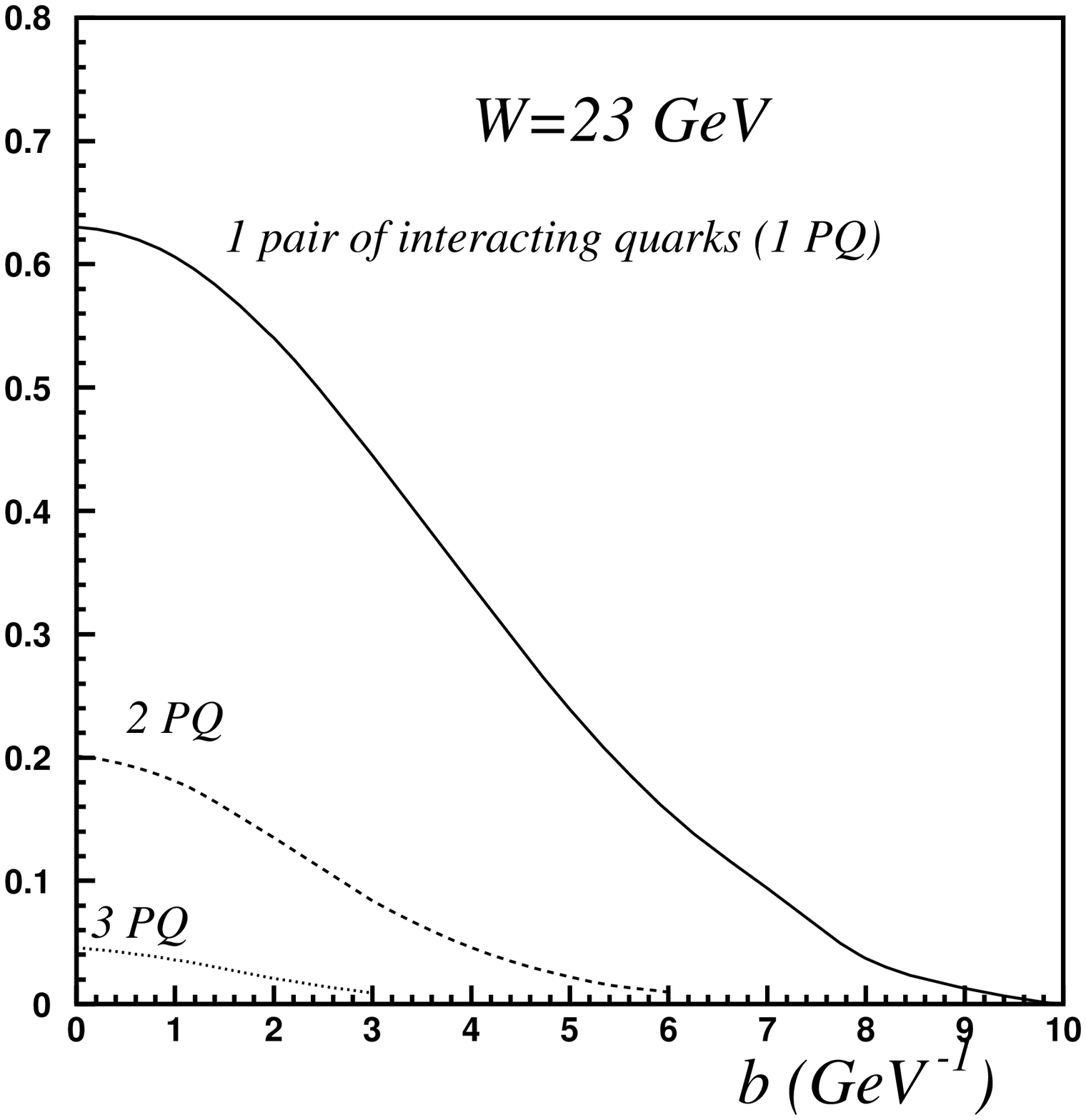,width=85mm} & 
\psfig{file=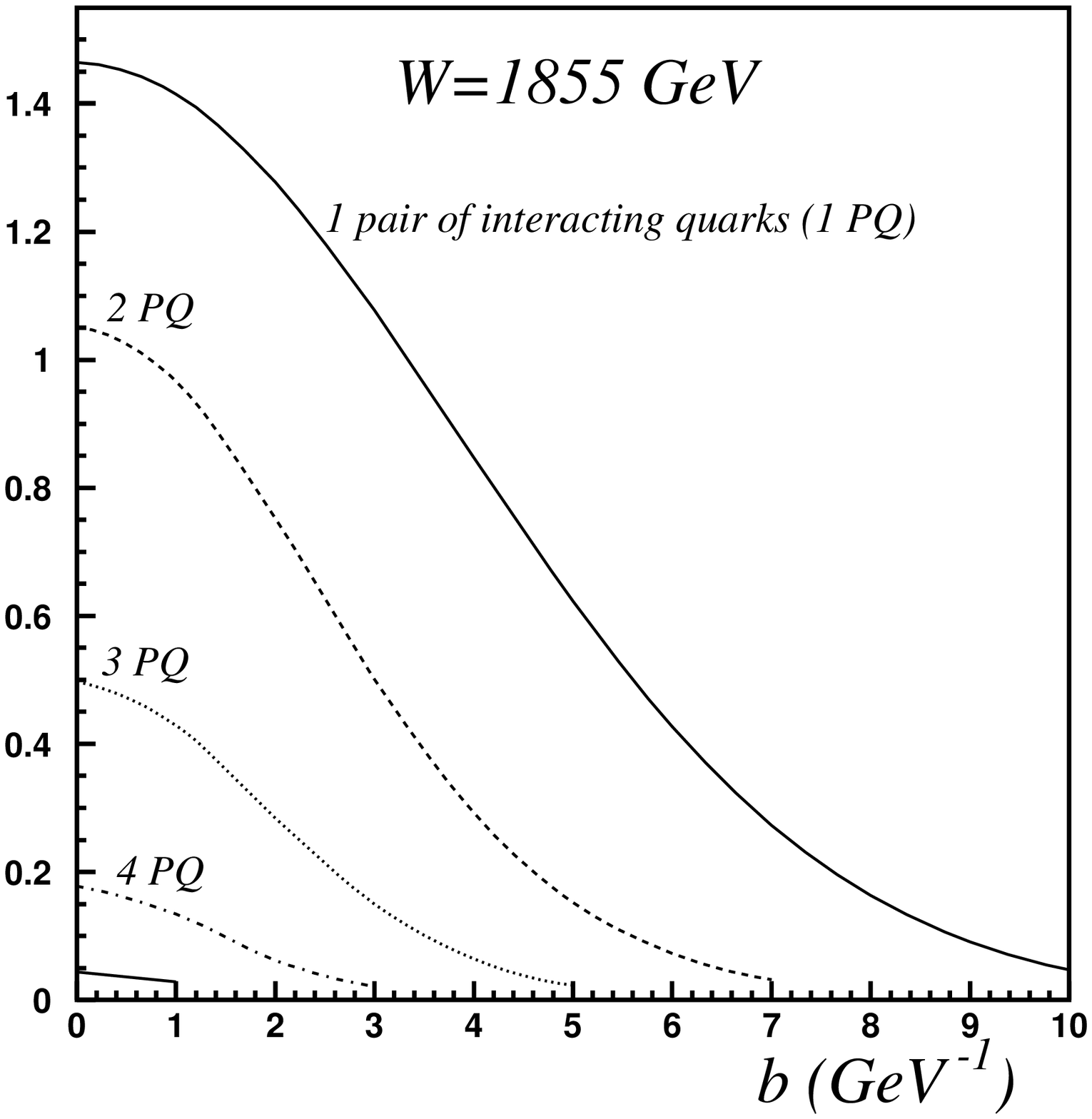,width=85mm}\\
       &  \\ 
\fig{Vklady}-a & \fig{Vklady}-b \\
 &  \\
\psfig{file=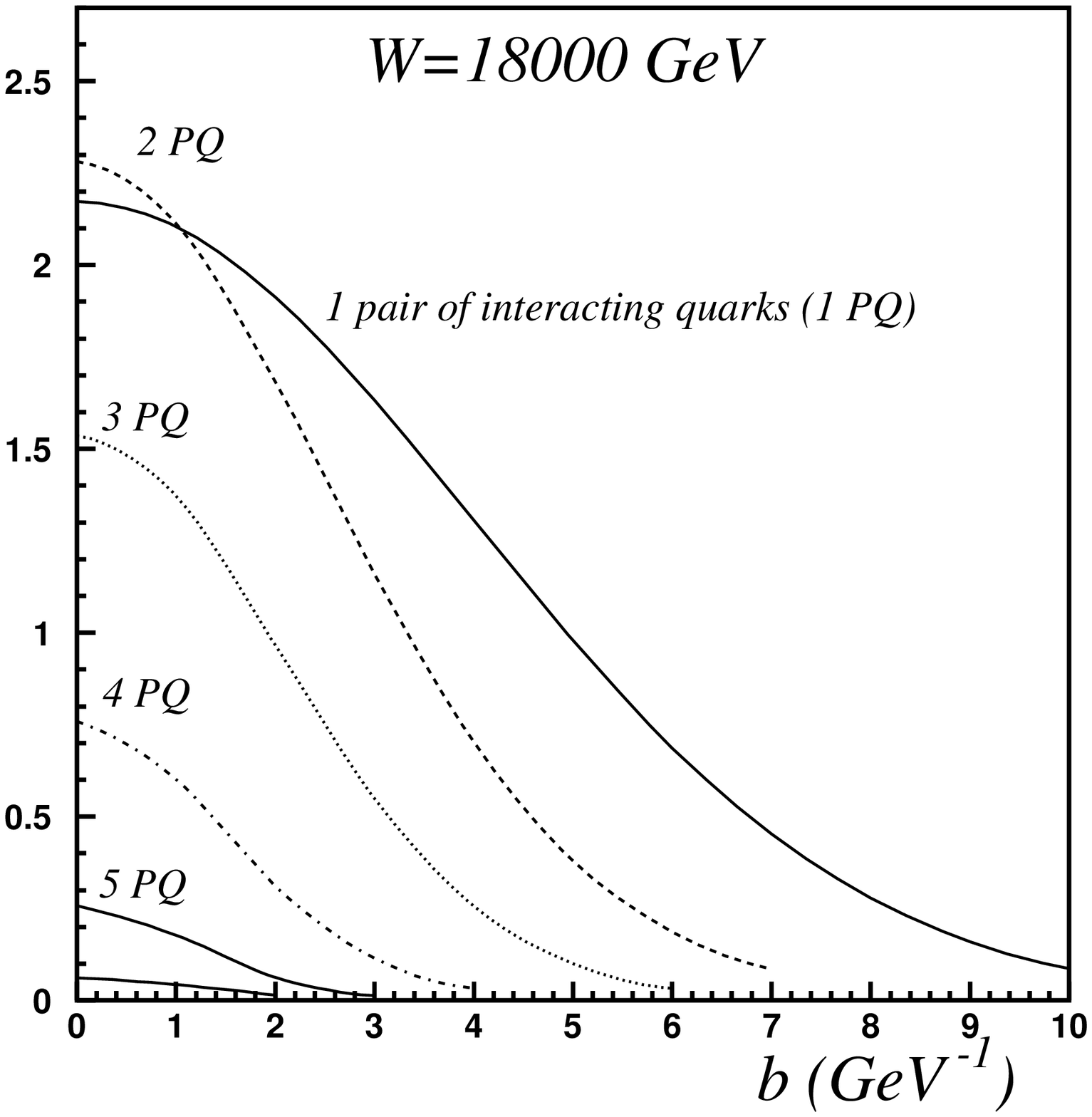,width=85mm}\\
       &  \\ 
\fig{Vklady}-c \\
 &  \\

\end{tabular}
\caption{\it The  
contributions to the elastic amplitude from
the  interactions of different numbers of quark pairs
at different energies.}
\label{Vklady}
\end{figure}

 The elastic amplitude in our  model has a sufficiently complex
structure. It
contains contributions
from  different configurations for  the  interactions
of quarks inside the protons.
We found all possible
terms which contribute to the elastic amplitude, see Appendix B,
and, therefore,  we can find out  
which  terms in elastic amplitude  are  
important at different  energies. It turns out, that even at low energies the 
two quark pairs interactions, i.e.
interactions  between two quarks in one proton
with two quarks in another proton, are not negligible. 
At Tevatron energy, $\,\sqrt{s}\,=\,1855\,GeV\,$, 
we  already need to take into account the
 contributions from the five pairs of interacting quarks. At
energy of order of the LHC energy,  $\,\sqrt{s}\,=\,15000\,GeV\,$, 
the contribution of the interaction of the seven quark pairs   
is valuable. At this energy the contribution
of   two quark pair interaction    exceeds  the
contribution
of one quark pair interaction, see Fig.~\ref{Vklady}. 
This is a signal that the parton clouds 
start to overlap, leading to the picture of \fig{aqmw}.

\begin{figure}[hptb]
\begin{center}
\psfig{file=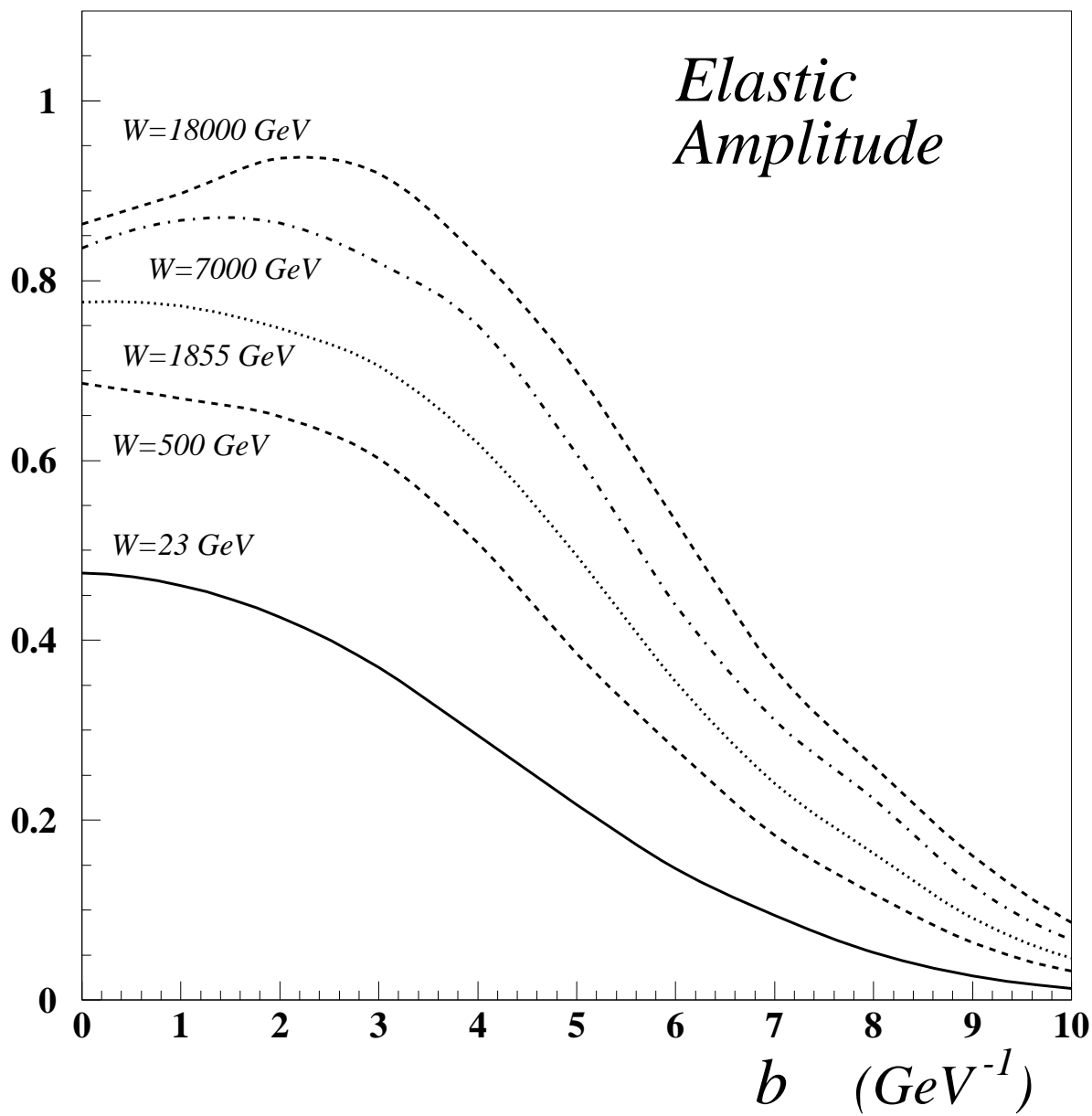,width=180mm}
\end{center} 
\caption{\it\large The elastic amplitude of the p-p scattering
as function of impact parameter at different energies. }
\label{Pr5}
\end{figure}

 From  Fig.~\ref{Vklady} we also see, that even at 
$\,\sqrt{s}\,=\,1855\,GeV\,$ energy, the one quark pair
contribution
to the elastic amplitude exceeds unity . 
The unitarization of the amplitude in this case is achieved not by
eikonalization of the quark-quark interaction 
but by including  more complicated configurations of the quarks inside of the 
protons in the 
elastic amplitude.
Therefore, the  form of impact parameter dependence
of the amplitude turns out to be  different from the 
usual Gaussian one. Indeed, the contribution of  two quark pair
interactions,
which have a negative sign in the amplitude 
is equal or larger than contribution of one quark pair interaction.
At the same time the one quark pair interaction is wider in the 
impact parameter space, see again  Fig.~\ref{Vklady}.
The contributions of all terms in  the elastic amplitude
lead, therefore, to the situation where the maximum of the
elastic amplitude moves from the zero impact parameter to the  
impact parameter
$\,b\,\approx\,2\,GeV^{-1}\,$, see Fig.~\ref{Pr5}.
This effect reflects very simple physics. At high energy
at small impact parameter the multi quark pair  interactions are 
important and elastic production becomes  mostly peripheral.

\section{Survival probability (SP) of the 'hard' processes
in p-p scattering}

In this section we consider the calculation of the survival 
probability for the process:$p + p 
\to p + [LRG] +  dijet +  [LRG] +  p$, 
where $p$ and $p$ are the colliding protons and LRG is 
the large rapidity gap.
In   this process there are
two  large rapidity gaps -- the intervals of rapidity without 
secondary hadrons.   The cross sections of such
processes are small in comparison with the inclusive production or
in other word, in comparison with the process where no rapidity 
gap is selected.  
The ratio of these two processes, exclusive and inclusive 
(or the process without LRG),
we call survival probability of large rapidity gap 
(see Ref.\cite{BJLRG}). In the simple case of the eikonal approach to proton-proton 
interaction, the survival probability is determined by a simple formula (see 
Ref.\cite{BJLRG,GLMLRG}):

\beq\label{SP}
\hat{S}^2\,=\,\frac{\int\,d^2\,b\,A(s,b)\,\sigma_{hard}(b)}
{\int\,d^2\,b\,\,\sigma_{hard}(b)\,}\,
\eeq
where 
\beq\label{SP1}
\,A(s,b)\,=\,e^{-\Omega(s,b)}\,
\eeq
is the probability that no inelastic interaction between
the scattered hadrons has happened at energy 
$\,\sqrt{s}\,$ and impact parameter $\,b\,$. 

Using a simple Gaussian parameterization for $\sigma_{hard}(b)$ , namely,
\beq\label{SP22}
\sigma_{hard}(b)\,=\,\sigma_{hard}\,\frac{e^{-\frac{b^2}{R^{2}_{H}}}}
{4\pi\,R^{2}_{H}}\,,
\eeq
with 
\beq \label{SP21}
R^2_H\,\,=\,\,8\,GeV^{-2}
\eeq
we reduce \eq{SP} to the form
\beq\label{SP11}
\hat{S}^2\,=\,\frac{1}{\pi\,R^{2}_{H}}\,\int\,d^2\,b\,A(s,b)\,
e^{-\frac{b^2}{R^{2}_{H}}}\,\,.
\eeq

\begin{figure}[hptb]
\begin{center}
\psfig{file=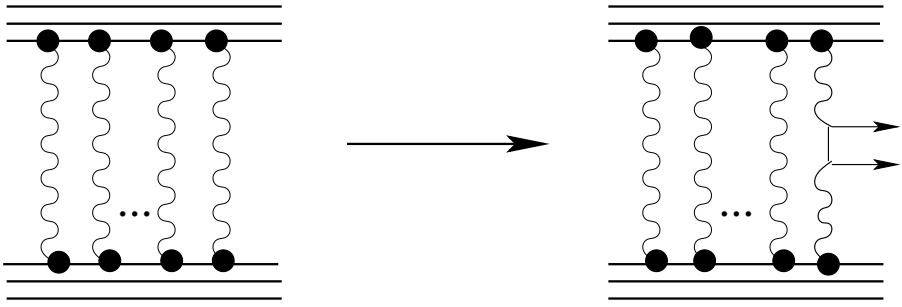,width=120mm}
\end{center} 
\caption{\it\large The first order amplitude of the LRG ''hard'' process. }
\label{SUP1}
\end{figure}

\begin{figure}[hptb]
\begin{center}
\psfig{file=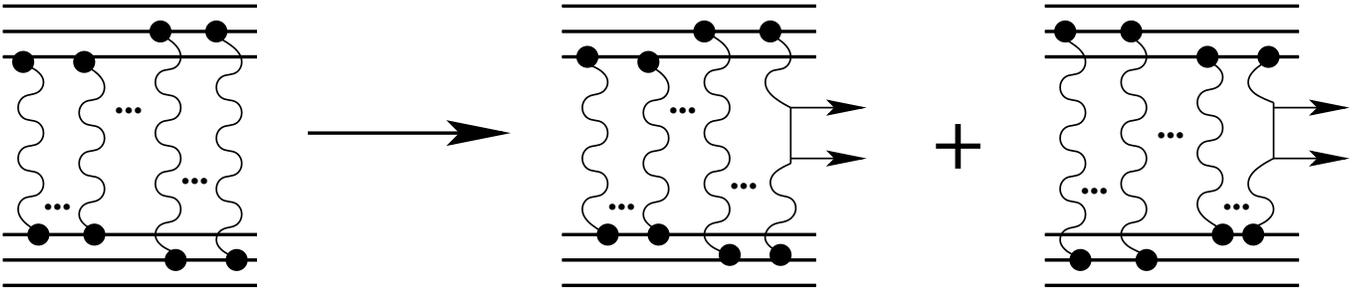,width=180mm}
\end{center} 
\caption{\it\large The general structure of the amplitude of the LRG ''hard'' 
process for the case of interaction of two quark pairs. }
\label{SUP2}
\end{figure}

 The difference of the calculation of the 
SP in  our model from the calculations above 
is that we consider as principle  degrees of freedom the constituent quarks but 
not  protons. Therefore, we need to determine  the expression for SP in  terms 
of the interacting quark pairs.
We begin with the  discussion of  the cross section  
for the exclusive ''hard'' production
in the case  of interaction of only  one pair of quarks. For this process
we have:

\beq\label{Sp5} \sigma_{hard}(b)\,=\,\sigma_{hard}\,
\Le\,\hat{A}_{1pair}(s,b)\Ra^2\,, \eeq with the amplitude
$\hat{A}_{1pair}(s,b)$, calculated from \eq{Ampli2} with he following
replacement

\beq\label{Sp7}
P_{q-q}(Y,b_{2})\,\rightarrow\,
\frac{e^{-\frac{b_{2}^2}{2\,R^{2}_{Q-H}}}}
{2\pi\,R^{2}_{H}}\hat{P}_{q-q}(Y,b_{2})\,,
\eeq
where 
\beq\label{Sp8}
\hat{P}_{q-q}(Y,b_{2})\,=\,
1\,-\,P_{q-q}(Y,b_{2})\,=\,e^{-\Omega_{q-q}(Y,b_{2})/2}\,
\eeq
in \eq{Sp7} means that t no inelastic interactions occur between 
the considered pair of quarks. This pair of quarks,
which produces ''hard'' jet, interacts elastically. 
The amplitude of this process is illustrated in  Fig.~\ref{SUP1}. 
In expression for the $\hat{A}(s,b)$,   
we need to introduce a new ''hard'' radius $\,R^{2}_{Q-H}\,$,
which related to the ''hard'' process on the quark level.
The numerical value of  $\,R^{2}_{Q-H}\,$ we  find in  the
following way. 
For  simplest process of ''hard''
production without
any  ''soft''  rescattering, 
the answer will be the same for any model :

\beq\label{Sp6}
e^{-\frac{b^2}{2\,R^{2}_{H}}}\,=\,
\,9\,\frac{9\,\alpha^2}{4\,\pi^2}\,
\int\,d^2\,b_{1}\,\int\,d^2\,b_{2}\,
e^{-\frac{3\alpha\,}{2}(\vec{b}-\vec{b}_{1})^2}\,
\,e^{-\frac{b_{2}^{2}}{2 R^{2}_{Q-H}}}\,
e^{-\frac{3\alpha}{2}\,(\vec{b}_{1}-\vec{b}_{2})^2}\,.
\eeq
Indeed, we can see, that  using  this equality  in
\eq{Sp5},  we  reproduce the expression given by \eq{SP22}.
From the \eq{Sp6}, we obtain the value of $\,R^{2}_{Q-H}\,$:

\beq\label{Sp666}
R^{2}_{Q-H}\,\approx\,1.1\,\,GeV^{-2}\,\,.
\eeq
Generalizing the procedure given by \eq{Sp7}
we obtain the answer for SP
amplitude $\hat{A}(s,b)$ in the all orders.  
Indeed,
any term of the amplitude given in Appendix B, see \eq{B1}, have
the integration over the  
product of the eikonalized amplitudes of the interacting quark pairs  
$\,P_{q-q}(b_i)\,...\,P_{q-q}(b_j)\,$.  
When we calculate the SP, we should  take into account the possibility that 
the ''hard'' process can occur in the interaction of any pair of quarks.
Therefore, for the pair with the ''hard''
process  the replacement of \eq{Sp7} should  be performed, whereas
other pairs of quarks will still interact elastically.
It means, that 
for any integral in expansion of \eq{B1} of elastic amplitude, we 
will make 
the following replacement for  any term of 
$\,P_{q-q}(b_i)\,...\,P_{q-q}(b_j)\,$ -  type
: 
\beq\label{Sp10}
P_{q-q}(b_i)\,...\,P_{q-q}(b_j)\,\rightarrow\,
\frac{e^{-\frac{b_{i}^{2}}{2\,R^{2}_{Q-H}}}}
{2\pi\,R^{2}_{H}}\hat{P}_{q-q}(b_i)\,...\,P_{q-q}(b_j)\,
\eeq
for each $\,P_{q-q}(b_k)\,$ in the chain 
$\,P_{q-q}(b_i)\,...\,P_{q-q}(b_j)\,$.
This procedure  is illustrated by  Fig.~\ref{SUP2} for the  
  case of the interaction of two
pairs of quarks. 

Performing these replacements and summing all terms,
we  obtain a new amplitude $\,\hat{A}(s,b)\,$ ( see  
\eq{B1}). In Appendix C
we show the example of this procedure for the 
$\,A_{3pairs}(s,b)\,$ term of the elastic amplitude $\,A(s,b)\,$.
With this amplitude, $\,\hat{A}(s,b)\,$, we determine  the 
''hard'' cross section as follows:
 
\beq\label{Sp100} 
\sigma_{hard}\,\propto\,\int\,d^2\,b\,
\Le\,\hat{A}(s,b)\Ra^2\,,
\eeq
where the first term of this expansion has the form: 

\beq\label{Sp111} 
\sigma_{hard}^{1}\,\propto\,
\frac{81}{4\,\pi^2\,R^{4}_{H}}
\int\,d^2\,b\,
\Le
\,\frac{9\,\alpha^2}{4\,\pi^2}
\int\,d^2\,b_{1}\,\int\,d^2\,b_{2}\,
e^{-\frac{3\alpha\,}{2}(\vec{b}-\vec{b}_{1})^2}\,
\hat{P}_{q-q}(Y,b_{2})\,e^{-\frac{b_{2}^{2}}{2 R^{2}_{Q-H}}}\,
e^{-\frac{3\alpha}{2}\,(\vec{b}_{1}-\vec{b}_{2})^2}\,\Ra^{2}
\eeq
Finally, the answer for the survival
probability factor in all orders
of the expansion of \eq{B1} looks as follows:

\beq\label{Sp12}
\hat{S}^{2}\,=\,4\,\pi^2\,R^{4}_{H}\,
\frac{\int\,d^2\,b\,\Le\,\hat{A}(s,b)\Ra^2\,}
{\int\,d^2\,b\,
\Le\,9\,
\,\frac{9\,\alpha^2}{4\,\pi^2}
\int\,d^2\,b_{1}\,\int\,d^2\,b_{2}\,
e^{-\frac{3\alpha\,}{2}(\vec{b}-\vec{b}_{1})^2}\,
\,e^{-\frac{b_{2}^{2}}{2 R^{2}_{Q-H}}}\,
e^{-\frac{3\alpha}{2}\,(\vec{b}_{1}-\vec{b}_{2})^2}\,\Ra^{2}}\,. 
\eeq
For example, there is  the first term, which contributes to the $\,\hat{S}^{2}\,$:

\beq\label{Sp13}
\hat{S}_{1}^{2}\,=\,
\frac{\int\,d^2\,b\,\Le\,
\,\frac{9\,\alpha^2}{4\,\pi^2}
\int\,d^2\,b_{1}\,\int\,d^2\,b_{2}\,
e^{-\frac{3\alpha\,}{2}(\vec{b}-\vec{b}_{1})^2}\,
\,e^{-\frac{b_{2}^{2}}{2 R^{2}_{Q-H}}}\,
\hat{P}_{q-q}(Y,b_{2})\,
e^{-\frac{3\alpha}{2}\,(\vec{b}_{1}-\vec{b}_{2})^2}\,
\Ra^2\,}
{\int\,d^2\,b\,
\Le
\,\frac{9\,\alpha^2}{4\,\pi^2}
\int\,d^2\,b_{1}\,\int\,d^2\,b_{2}\,
e^{-\frac{3\alpha\,}{2}(\vec{b}-\vec{b}_{1})^2}\,
\,e^{-\frac{b_{2}^{2}}{2 R^{2}_{Q-H}}}\,
e^{-\frac{3\alpha}{2}\,(\vec{b}_{1}-\vec{b}_{2})^2}\,\Ra^{2}}\,. 
\eeq

The values of  the  survival probability factor, $\,\hat{S}^{2}\,$,
calculated in this approach,
are shown in the Table~\ref{SP66}. 
\begin{table}
\begin{tabular}{|c|c|c|c|}
\hline
\, & \, & \,&\,\\ 
$ SP\,(\,\sqrt{s}=23\,GeV )$ &
$ SP\,(\,\sqrt{s}=1855\,GeV $ & 
$ SP\,(\,\sqrt{s}=15000\,GeV ) $ & 
$ SP\,(\,\sqrt{s}=18000\,GeV )$ \\
\, & \, & \,&\,\\
\hline
\, & \, & \,&\,\\
0.23 & 0.052 & 0.0175 & 0.0155 \\
\, &  \, & \, &\,\\
\hline
\end{tabular}
\caption{\it The survival probability factor
as a function of $\sqrt{s}$}
\label{SP66}
\end{table}
We see, that these values  are close  
to the
values of survival probability for the  central 
diffraction process calculated  in Ref. \cite{KMR}.
\begin{figure}[hptb]
\begin{tabular}{ c c}
\psfig{file=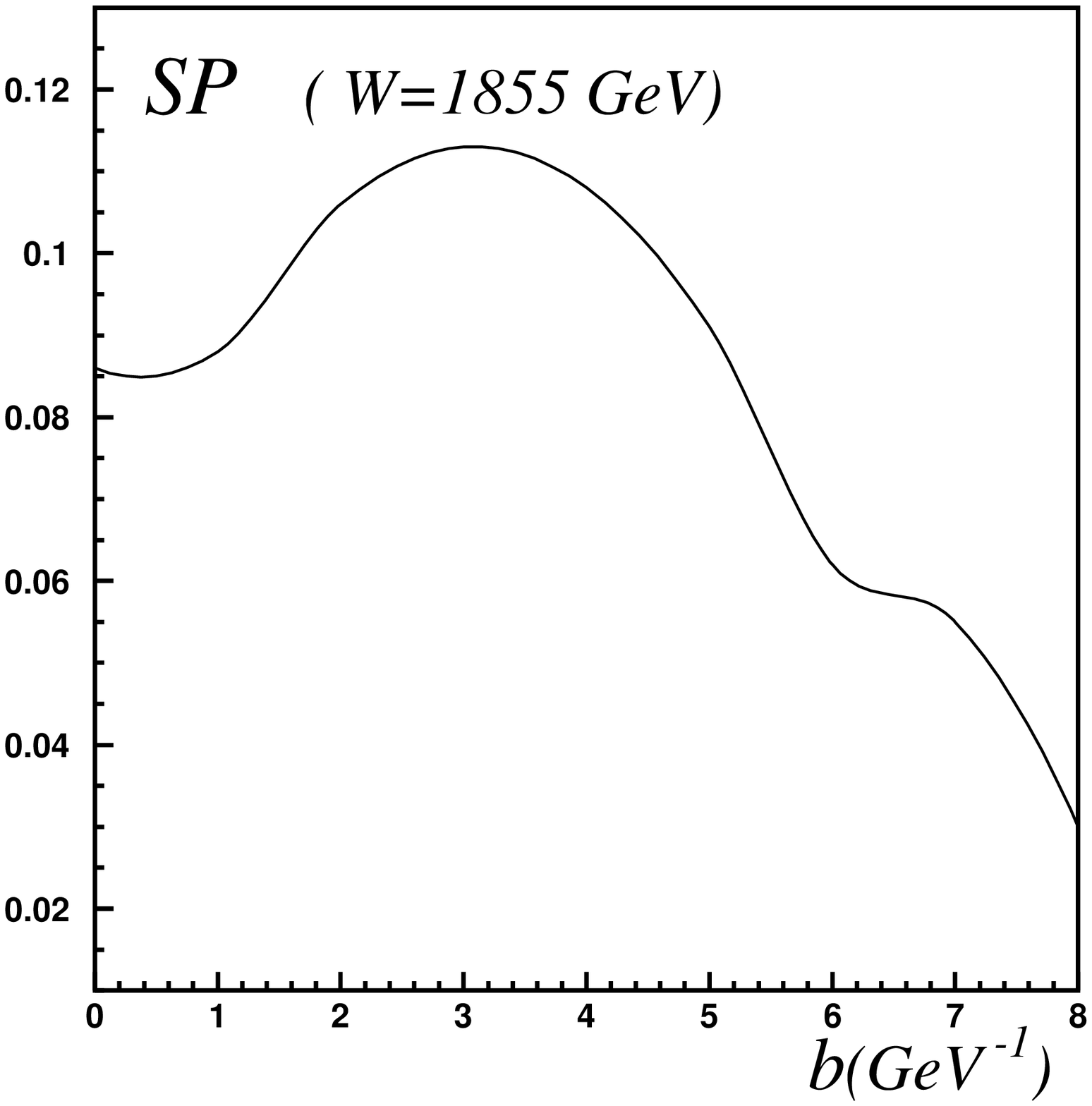,width=85mm} & 
\psfig{file=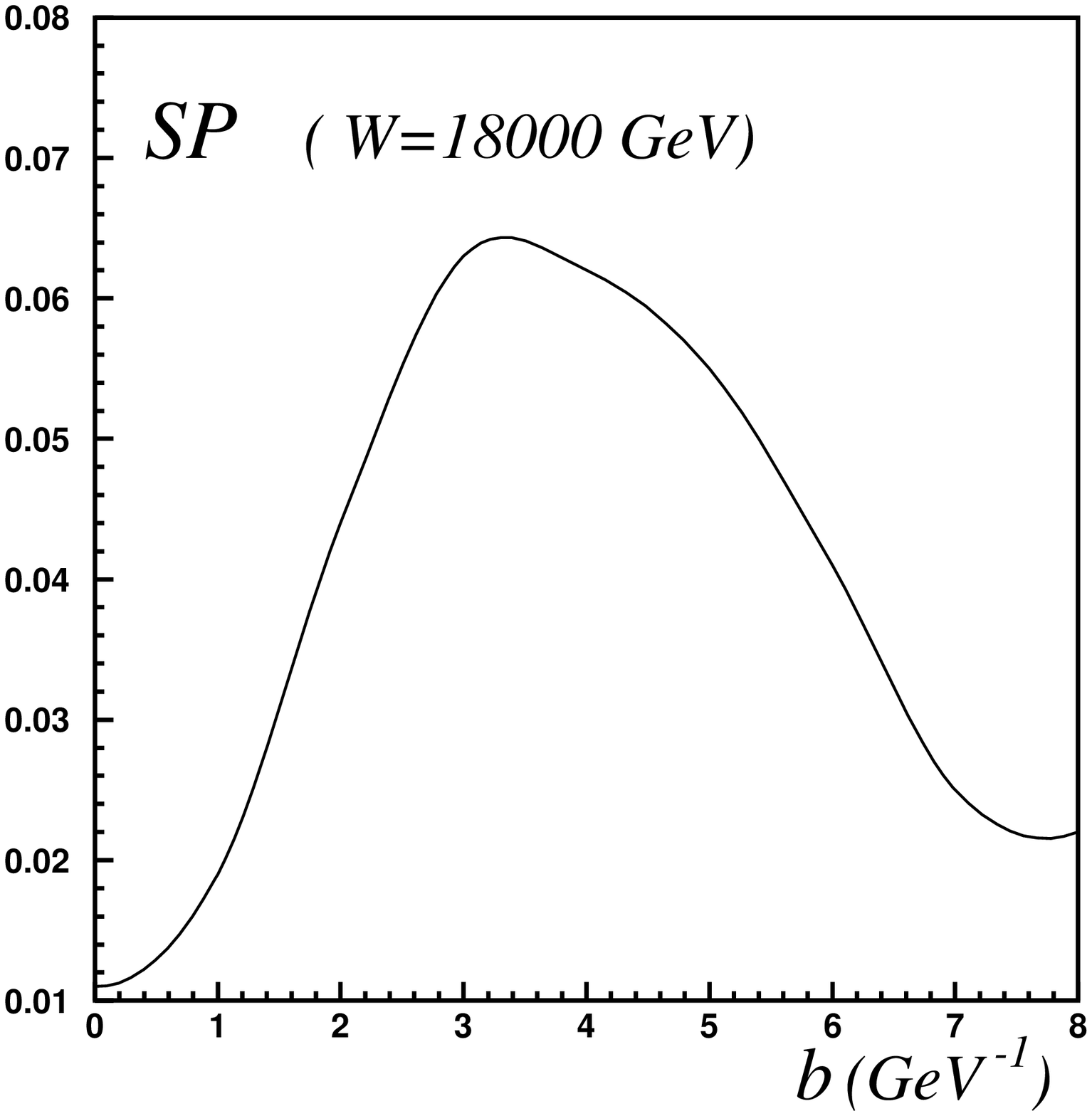,width=85mm}\\
       &  \\ 
\fig{SurProb}-a & \fig{SurProb}-b \\
 &  \\
\end{tabular}
\caption{\it The impact parameter behavior of the 
integrant in the numerator of  \eq{Sp12} 
at energies $\,W=1855\,\,GeV\,$ and  $\,W=18000\,\,GeV\,$.}
\label{SurProb}
\end{figure}
We also performed the calculation of the integrant in the
numerator of  \eq{Sp12} , which  is  proportional to the amplitude
of the ''hard'' central diffraction process. The plots of this integrant
as a function of impact parameter at different energies are presented
in  Fig.~\ref{SurProb}.
Considering these graphs we conclude, that the main contribution to the amplitude
of the central diffraction ''hard'' production
comes from the non-central values of  impact parameter  and  it is almost zero
for the central impact parameter at high energies.

\section{The cross section of the diffraction dissociation }

\begin{figure}[hptb]
\begin{tabular}{ c c}
\psfig{file=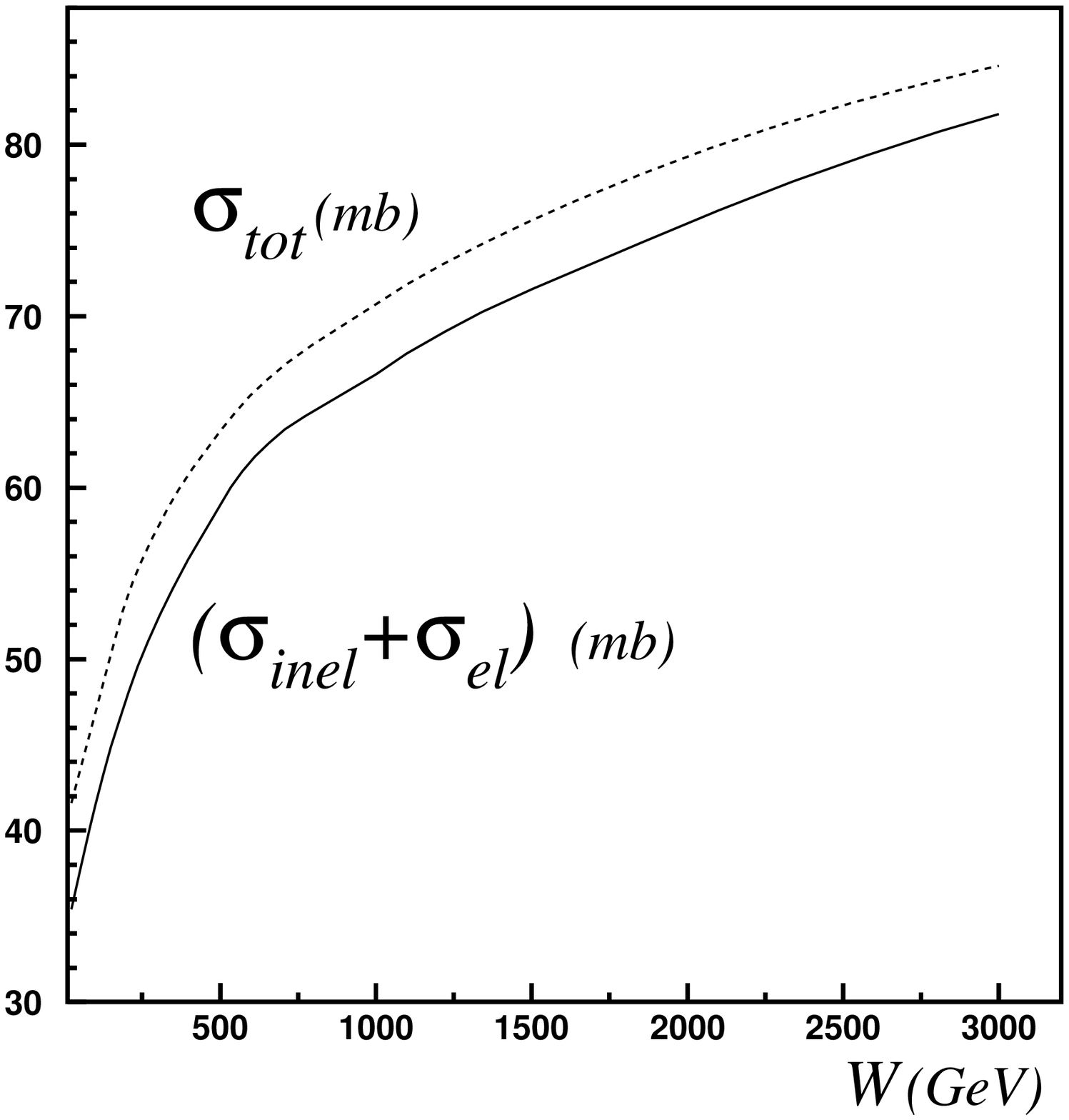,width=85mm} & 
\psfig{file=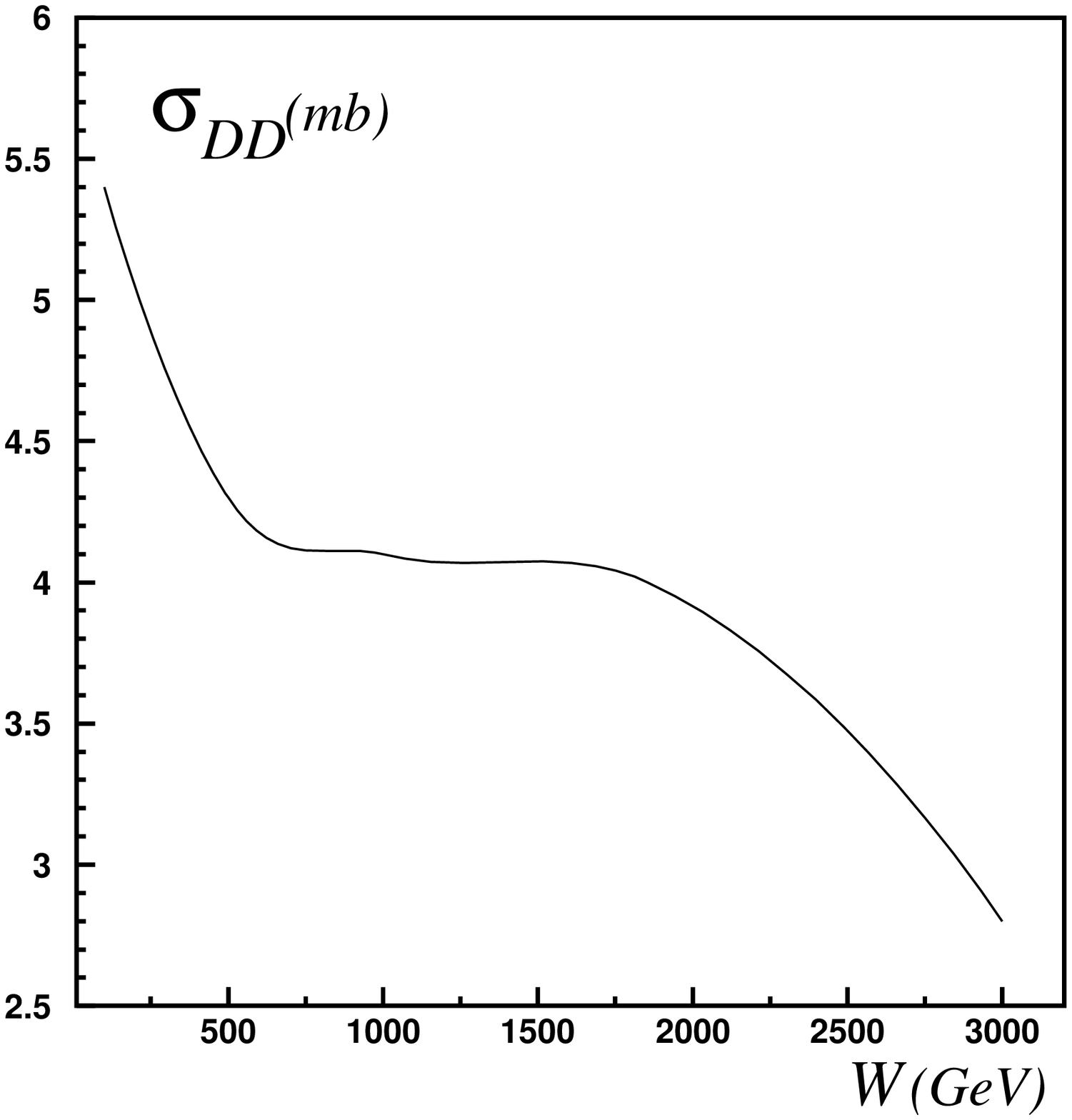,width=85mm}\\
       &  \\ 
\fig{Vkladydd}-a & \fig{Vkladydd}-b \\
 &  \\

\end{tabular}
\caption{\it 
The sum of elastic and inelastic cross 
sections in comparison with the total cross section and DD cross section
as the function of $W\,=\,\sqrt{s}$.}
\label{Vkladydd}
\end{figure}

   The elastic amplitude of the considered  model 
accounts many Pomeron exchanges between different
pairs of quarks. Now,  we 
want to  calculate the cross section of the 
diffraction dissociation processes taking into account all these processes. 
The diffractive 
dissociation (DD)
processes for us are  all processes where
no particle have been produced in the rapidity region between two
scattered protons. From the unitarity constraint we have:

\beq\label{NDD}
\sigma_{tot}\,=\,\sigma_{el}\,+\,\sigma_{inel}\,+\,\sigma_{DD}\,.
\eeq
Therefore:

\beq\label{NDD1}
\,\sigma_{DD}\,=\,\sigma_{tot}\,-\,\sigma_{el}\,-\,\sigma_{inel}.
\eeq
From our previous calculations we know the values of 
$\,\sigma_{el}\,$ and $\,\sigma_{tot}\,$. So, in order to calculate
$\,\sigma_{DD}\,$ we need to calculate the value of $\,\sigma_{inel}\,$.
The calculation of $\,\sigma_{inel}\,$ is pretty simple.
In the first order, where

\beq\label{DD1}
\sigma_{total}^{1}\,=\,2\,\int\,d^2\,b\,A_{1pair}(s,b)\,=\,
18\,\int\,d^2\,b\,P_{q-q}(Y,b)\,,
\eeq
for $\,\sigma_{inel}\,$ we have:

\beq\label{NDD2}
\sigma_{inel}\,=\,\int\,d^2\,b\,A_{1pair}^{inel}(s,b)\,=\,
9\,\int\,d^2\,b\,P_{q-q}^{inel}(Y,b)\,,
\eeq
with

\beq\label{DD6}
P^{inel}_{q-q}(Y,b)\,=\,1\,-\,e^{-\Omega_{q-q}(Y,b)}\,.
\eeq
The calculation of $\,\sigma_{inel}\,$ for all
possible interactions between the pairs of the quarks in the protons 
we perform using the following simple receipt. In each expression for
$\,\sigma_{tot}\,$  

\beq\label{NDD3}
\sigma_{total}\,=\,2\,\int\,d^2\,b\,A(s,b)\,,  
\eeq
we make replacement:

\beq\label{NDD33}
P_{q-q}(Y,b_i)... P_{q-q}(Y,b_j)\,\rightarrow\,
\frac{1}{2}\,P_{q-q}^{inel}(Y,b_i)... P_{q-q}^{inel}(Y,b_j)\,,
\eeq
with the $\,P_{q-q}^{inel}\,$ given by \eq{DD6}.
After the calculations of $\,\sigma_{inel}\,$ with the help of \eq{NDD2}
we obtain the value of $\,\sigma_{DD}\,$.
The result of calculations for the energy range
$\,W\,=\,100\,-\,3000\,\,\,GeV\,$ is shown in  Fig.~\ref{Vkladydd},
as well as the sum of elastic and inelastic cross 
sections in comparison with the total cross section.

For the energies higher  than $\,W\,=\,3000\,\,GeV\,\, $,  the cross section
of diffractive dissociation process is almost zero, see 
Table~\ref{DD2}, as it must be at high energies when we approach
the ''black disc'' limit.
Nevertheless, it is important to notice, that in calculations of DD processes, 
we neglected the  diffraction dissociation processes with the  large 
mass production. Such processes may be 
described if we introduce in the model  the triple Pomeron vertex, see
\cite{BLN}. We did not consider this vertex in our calculations, 
therefore, our result for the cross section of the DD processes,
which is the sum of single diffraction and double diffraction processes, is smaller
than it could be in the case when the triple Pomeron vertex is included.

\begin{table}
\begin{tabular}{|c|c|c|c|}
\hline
\, & \, & \,& \\ 
$ \sigma_{DD}(\sqrt{s}=7000\,GeV ) $ & 
$ \sigma_{DD}(\sqrt{s}=10000\,GeV ) $ & 
$ \sigma_{DD}(\sqrt{s}=15000\,GeV ) $ & 
$ \sigma_{DD}(\sqrt{s}=18000\,GeV ) $\\
\, & \, & \,& \\
\hline
\, & \, & \,& \\
2.3\,\,mb & 0.4\,\,mb & 0.35\,\,mb & 0.3\,\,mb \\
\, &  \, & \, & \\
\hline
\end{tabular}
\caption{\it The diffractive dissociation cross section
at high energies as a function of $\sqrt{s}$.}
\label{DD2}
\end{table}

\section{Conclusion}
 
 In this paper we consider  the proton-proton interactions
in the framework of the Constituent Quark Model (CQM) at high energies.
The typical ''soft'' process of p-p scattering we described taking    
into account all 
interactions between pairs of the quarks in the protons,
modeling the quark-quark interaction by eikonal formula. 
It turns out , that in this model
the interactions of one, two or three quarks from one proton
with  two or three quarks from the second proton surprisingly
become  essential  even at low energies. 
Indeed, if 
we look at Fig.~\ref{Vklady}, we  see, that even
at energy $\,\sqrt{s}\,=23\,GeV\,$ the contribution
from the interaction of two pairs of quarks is  approximately one third
from the contribution of  one pair.
Of course, at higher energies the contributions of more  
interacting quark pairs became to be more important. 
For example, at LHC 
energy the interactions of  one and two pairs of quarks
equally contribute to the elastic amplitude
and at this energy we need to account the  contribution of seven 
quark pairs.
This interaction picture leads 
to the very natural scenario of the unitarization. 
At high energies the 
contributions from the interaction of one and two  pairs of quarks
cancel
each other, and final amplitude at given impact parameter is smaller than one,
see Fig.~\ref{Pr5}. We see, that in this case
the unitarization is achieved
by exploring internal structure of the proton rather 
than details of interactions. However,
we should stress, that we assume that the quark-quark interactions 
is described by the eikonal 
approach. This specific mechanism manifests itself in
the form of the b-dependence of the
amplitude, which  is quite different from the usual $\,b\,$  dependence.
Our amplitude has maximum 
at $\,b\,\approx\,2\,GeV^{-1}\,$ at high energy, that means
that elastic production is mostly peripheral at high energies.

  Another result , observed in the present model and  related
to the unitarization of the amplitude,
is the relatively small value of the amplitude at
zero impact parameter. Indeed, 
even at energy $\,\sqrt{s}\,=18000\,GeV\,$ at $\,b\,=\,0\,$
the amplitude is not close to one. The interactions are still 
''grey'' and not ''black''.  Many Pomeron interactions
between different quark configurations in both protons lead
to the ''grey '' picture in spite of the large
contributions of the one, two or  three interacting quark pairs.    

  The considered model fits the experimental data pretty 
well, see Fig.~\ref{Pr3}. Using the parameters of  the ''input''  Pomeron
, which we obtained through the data fitting, we can 
predict the values of the cross sections at high energies.
Doing so, we did not find the deviation of the behavior of the elastic
slope, $\,B_{el}\,$, from the simple linear parameterization, see plot of 
Fig.~\ref{Pr4}. The explanation of
this effect is the following. In spite of  the hope, that the constituent
quarks will have strong overlapping at high energies  and this effect will change
the energy behavior of the $\,B_{el}\,$, it actually does not happen.
The value of the size of  constituent quark obtained from the data fit  is small, 
$\,R^{2}_{quark}\,\approx\,0.16\, GeV^{-2}\,\,$. The slope
of the initial quark-quark amplitude, which is our ''input'' Pomeron,
also turns out to be small , $\alpha^{'}\,\approx\,0.08\, GeV^{-2}\,$.
Therefore, even at energy  $\,\sqrt{s}\,=18000\,GeV\,$ we obtain for 
the radius of the 
constituent quark $\,R_{quark}\,\approx\,0.3\,-\,0.4\, fm\,\,$, that is still
in two times smaller than the radius of the proton. 
This again supports the conclusion, that
our interactions at LHC energy still far from the ''black'' disk limit. 

 We also calculated the survival probability factor
for different energies and
it turns out to be similar to  the values obtained in model proposed in Ref.
\cite{KMR}. The parameters of the ''input'' Pomeron
in our paper and in Ref. \cite{KMR} are close, in spite of the fact that we did not 
involve any non-perturbative physics in our calculations. 
The intercept of the ''input'' Pomeron, which we obtained ,
is $\,\Delta\,\approx\,0.12\,$ . This intercept 
is larger than the intercept of the 
paper \cite{KMR}. The  slope,  
$\alpha^{'}\,\approx\,0.08\, GeV^{-2}\,$,
is very close to the one of the paper \cite{KMR}.
Considering the impact parameter dependence of the 
amplitude of  the exclusive central diffraction
''hard'' process, see Fig.~\ref{SurProb}, we see, that the maximum
of the amplitude locates at peripheral impact parameter, 
$\,b\,\approx\,3\,GeV^{-2}\,$ at energy of LHC.
The probability of this process at zero impact parameter is almost zero. 

 The values of the parameters  of our ''input'' Pomeron, 
the slope  ($\alpha^{'}_P\,\approx\,0.08 GeV^{-2}$ and the intercept 
$\Delta \approx \,0.2$, give rise to the idea that  
this Pomeron is not ''soft'' (see Ref. \cite{POM} for the typical parameters of the soft 
Pomeron). The parameters that we obtained related to so called ''hard" Pomeron. 
Therefore, the 
one of the result of this paper is the idea that, perhaps, we do not need to introduce 
''soft" 
Pomeron in order to describe the ''soft'' data.
Considering the internal structure of the colliding hadrons as well 
as the unitarization corrections for the amplitude we can describe p-p
data using only one, ''hard'' Pomeron.

\section*{Acknowledgments:}
We want to thank Asher Gotsman,  Uri Maor and Alex Prygarin for very useful
discussions on the subject
of this paper. This research was supported in part  by the Israel Science Foundation,
founded by the Israeli Academy of Science and Humanities and by BSF grant \# 20004019.
The research of S.B. was supported by Minerva Israeli-Germany Science foundation.
One from the authors, S.B., want also to thank the II Theoretical Physics
Institute, DESY for hospitality and support.

\newpage
\appendix 

\section*{Appendix A:}

 \renewcommand{\theequation}{A.\arabic{equation}}
\setcounter{equation}{0}

\begin{figure}[hptb]
\begin{center}
\psfig{file=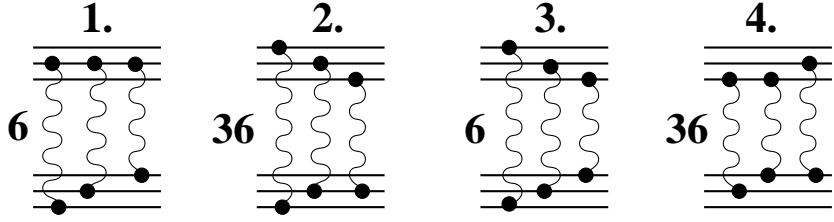,width=110mm}
\end{center} 
\caption{\it\large Four diagrams of the three quark pair  interaction. 
The wave line describes the 
eikonal amplitude for quark-quark interaction $P(Y,b)$.}
\label{App3}
\end{figure}

\begin{figure}[hptb]
\begin{center}
\psfig{file=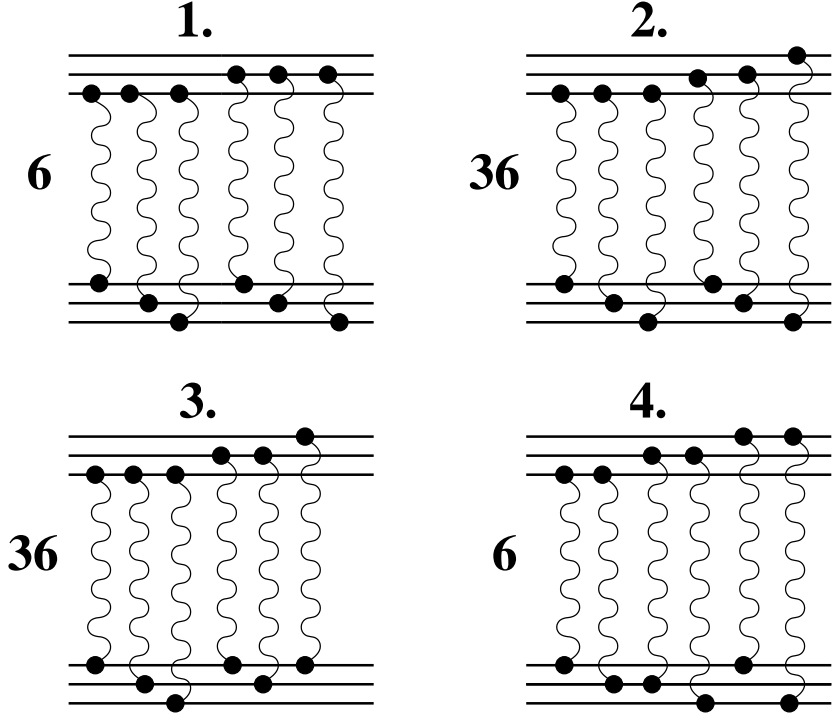,width=110mm}
\end{center} 
\caption{\it\large Four diagrams of the six quark pairs interactions.}
\label{App6}
\end{figure}

 In this appendix, as an  example of our calculations,  we calculate two 
diagrams, which contribute
to the elastic amplitude: the first diagram of  
Fig.~\ref{App3} and the first diagram of  Fig.~\ref{App6}.

 We begin with  the first diagram of Fig.~\ref{App3}. There are
six  diagrams of this type and we need two vertices (see \eq{Vert1} and 
\eq{Vert3}) for the calculations:
$$
V_{up}(k)=e^{-k^2/(6\,\alpha)}\,,
$$
and
$$
V_{down}(k)=e^{-\frac{1}{8\,\alpha}(\vec{q}_{2}-\vec{q}_{3})^2-
\frac{1}{6\,\alpha}(\vec{q}_{1}-\vec{q}_{2}/2-\vec{q}_{3}/2)^2\,}\,,
$$
where $\,q_i\,$ is the momentum transferred of each single Pomeron,
and $\,k\,\,=q_1+q_2+q_3\,$.
So, we have for our diagram:

\beq\label{3Pom}
D_{1,3pairs}(k)\,= \,6\,\int\,\prod_{i=1}^{3}\,\frac{d^2\,q_i}{4\,\pi^2}\,
P_{q-q}(q_i)\,e^{-k^2/(6\,\alpha)}\,
e^{-\frac{1}{8\,\alpha}(\vec{q}_{2}-\vec{q}_{3})^2-
\frac{3}{8\,\alpha}(\vec{q}_{1}-\vec{k}/3)^2\,}\,(4\,\pi^2\,
\delta^{2}(\vec{k}-\vec{q}_{1}-\vec{q}_{2}-\vec{q}_{3})\,)\,
\eeq
where $\,P_{q-q}(q_i)\,$ is the 
Fourier transform of amplitude given  by \eq{EikPom}.

 Since  we have no simple analytic expression
for $\,P_{q-q}(q_i)\,$, 
we make Fourier transform for
each function  $\,P_{q-q}(q_i)\,$ and  rewrite our expression
in  terms of  $\,P_{q-q}(b_i)\,$:
\beq
P_{q-q}(q_i)\,=\,\int\,d^2\,b_i\,P_{q-q}(b_i)\,
e^{i\vec{q}_i\,\vec{b}_i}\,.
\eeq
Here
$\,b_i\,$ is the impact parameter variable, which is conjugated to 
 momentum $\,q_i\,$ of the  single Pomeron. We also  make Fourier transform 
from $\,k\,$ to impact parameter variable $\,b\,$.
We obtain:

\begin{eqnarray}\label{3Pom2}
D_{1,3pairs}(b)\,& = &\,\frac{6}{64\,\pi^6}\,
\int\,\prod_{i=1}^{3}\,d^2\,b_i\,P_{q-q}(b_i)\,
\int\,\prod_{i=1}^{3}\,d^2\,q_i\,
\,\int\,d^2\,k\,\,\,e^{-i\,\vec{k}\,\vec{b}}\\
\,&\,&\,
e^{i\,\sum_{i=1}^{3}\,\vec{q}_{i}\,\vec{b}_{i}}\,
e^{-k^2/(6\,\alpha)}\,
e^{-\frac{1}{8\,\alpha}(\vec{q}_{2}-\vec{q}_{3})^2-
\frac{3}{8\,\alpha}(\vec{q}_{1}-\vec{k}/3)^2\,}\,\,
\delta^{2}(\vec{k}-\vec{q}_{1}-\vec{q}_{2}-\vec{q}_{3})\,.
\end{eqnarray}

 To make this expression easier for calculations we
introduce new variables:

\beq\label{3Pom3}
x_1=q_2-q_3\,;\,\,x_2=q_1-k/3\,;\,\,x_3=q_3\,\,,\\
\eeq
$$
q_1=x_2+k/3\,;\,\,q_2=x_1+x_3\,;\,\,q_3=x_3\,\,.
$$
The Jacobian of this transformations is equal to unity;  and we obtain:

\begin{eqnarray}\label{3Pom4}
D_{1,3pairs}(b)\,& = &\,
\frac{6}{64\,\pi^6}\,
\int\,\prod_{i=1}^{3}\,d^2\,b_i\,P_{q-q}(b_i)\,
\int\,\prod_{i=1}^{3}\,d^2\,x_i\,
\,\int\,d^2\,k\,\,\,e^{-i\,\vec{k}\,\vec{b}}\,
e^{i\,\vec{b}_{1}\,(\vec{x}_{2}+\vec{k}/3)}\,\\
\,&\,&\,
e^{i\,\vec{b}_{2}\,(\vec{x}_{3}+\vec{x}_{1})}\,
e^{i\,\vec{b}_{3}\,\vec{x}_{3}}\,\,
e^{-\frac{1}{6\,\alpha}\,k^2\,-\,
\frac{1}{8\,\alpha}\,x_{1}^{2}\,-\,
\frac{3}{8\,\alpha}\,x_{2}^{2}\,}\,\,
\delta^{2}(2\vec{k}/3-\vec{x}_{1}-\vec{x}_{2}-2\vec{x}_{3})\,.
\end{eqnarray}
Performing  the $\,x_3\,$ integration we have:

\begin{eqnarray}\label{3Pom5}
D_{1,3pairs}(b)\,& = &\,
\frac{6}{4\,64\,\pi^6}\,
\int\,\prod_{i=1}^{3}\,d^2\,b_i\,P_{q-q}(b_i)\,
\int\,\prod_{i=1}^{2}\,d^2\,x_i\,
\,\int\,d^2\,k\,\,
\,e^{i\,\vec{k}\,(-\vec{b}+\vec{b}_{1}/3+
\vec{b}_{2}/3+\vec{b}_{3}/3)}\,\\
\,&\,&\,
e^{i\,\vec{x}_{1}\,(\vec{b}_{2}/2-\vec{b}_{3}/2)}\,
e^{i\,\vec{x}_{2}\,(\vec{b}_{1}-\vec{b}_{2}/2-
\vec{b}_{3}/2)}\,
e^{-\frac{1}{6\,\alpha}\,k^2\,-\,
\frac{1}{8\,\alpha}\,x_{1}^{2}\,-\,
\frac{3}{8\,\alpha}\,x_{2}^{2}\,}\,.
\end{eqnarray}
Performing one integration over
$\,x_i\,$ and $\,k\,$, we obtain the final answer for
this diagram:

\beq\label{3Pom55}
D_{1,3pair}(b)\,= \,
\frac{6\,\alpha^3}{2\,\pi^3}\,
\int\,\prod_{i=1}^{3}\,d^2\,b_i\,P_{q-q}(b_i)\,
e^{-\frac{3\,\alpha}{2}\,(\vec{b}-\vec{b}_{1}/3-
\vec{b}_{2}/3-\vec{b}_{3}/3\,)^2\,-\,\frac{\alpha}{2}\,
(\vec{b}_{2}-\vec{b}_{3})^2\,-\,
\frac{2\,\alpha}{3}\,(\vec{b}_{1}-\vec{b}_{2}/2-
\vec{b}_{3}/2)^2\,} 
\eeq

  As the second example of the 
calculations technique  we  calculate the first diagram of 
Fig.~\ref{App6}. The main steps in calculation
of this diagram are the same as in the calculation
of the first diagram of Fig.~\ref{App3}. Therefore,
now we are not  focused on the detailed explanations
of main steps, but we rather  clarify the principal points
of the calculations technique.  

Two  vertices of the Pomerons-quarks coupling have
the following form 
(see \eq{Vert2} and \eq{Vert3}):

$$
V_{up}(k)=e^{-\frac{1}{6\,\alpha}\,(\vec{q}_1+
\vec{q}_2+\vec{q}_3)^2-\frac{1}{6\,\alpha}\,
(\vec{q}_4+\vec{q}_5+\vec{q}_6)^2\,+
\frac{1}{6\,\alpha}\,(\vec{q}_1+
\vec{q}_2+\vec{q}_3)\,(\vec{q}_4+\vec{q}_5+\vec{q}_6)}\,,
$$
and
$$
V_{down}(k)=e^{-\frac{1}{8\,\alpha}(\vec{q}_{1}+\vec{q}_{4}-
\vec{q}_2-\vec{q}_5)^2-
\frac{1}{6\,\alpha}(\vec{q}_{3}+\vec{q}_6-
\vec{q}_{1}/2-\vec{q}_{2}/2-\vec{q}_{4}/2-
\vec{q}_{5}/2)^2/,}\,.
$$
Here, the total momentum
transferred is $\,k\,=\,\sum_{i=1}^{6}\,q_{i}$.
We  change the  variables, in order to obtain
a simple Gaussian expression for these vertices:

\begin{eqnarray}\label{6Pom1}
x_1-x_2=q_1+q_2+q_3\,&\,&\,x_1+x_2=q_4+q_5+q_6\,;\\
x_3=q_1+q_4-q_2-q_5\,&\,&\,x_4=3q_3/2+3q_6/2-k/2\,;\\
x_5=q_5\,&\,&\,x_6=q_6\,.
\end{eqnarray}
And for old variables $\,q_i\,$ and $\,k\,$ we have:

\begin{eqnarray}\label{6Pom2}
k=2\,x_1\,&\,&\,q_1=-x_1/3-x_2+x_3/2-x_4/3+x_5+x_6\,;\\
q_2=2x_1/3-x_3/2-x-4/3-x_5\,&\,&\,q_3=2x_4/3-x_6+2x_1/3\,;\\
q_4=x_1+x_2-x_5-x_6\,&\,&\,q_5=x_5\,\,q_6=x_6\,\,.   
\end{eqnarray}
Here momenta $\,q_1\,$ are not independent. For these momenta we
used the delta function constraint:

$$
q_1=k\,-\,q_2\,-\,q_3\,-\,q_4\,-\,q_5\,-\,q_6\,.
$$
Using \eq{6Pom2} we calculate the Jacobian of the
variable change, it is equal to  $\,\frac{4}{9}\,$. 
In new variables the vertices look very simple:

\beq\label{6Pom3}
V_{up}(k)=e^{-\frac{1}{6\,\alpha}\,x_{1}^{2}-
\frac{1}{2\,\alpha}\,x_{2}^{2}}\,,
\eeq
and

\beq\label{6Pom4}
V_{down}(k)=e^{-\frac{1}{8\,\alpha}\,x_{3}^{2}-
\frac{1}{6\,\alpha}\,x_{4}^{2}}\,.
\eeq

  We  consider the exponents that  stem from the Fourier transform
from momentum to impact parameter representation:

\beq\label{6Pom5}
e^{-i\,\vec{k}\,\vec{b}}\,
e^{i\,\prod_{i=1}^{6}\,\vec{q}_{i}\,\vec{b}_{i}}\,\,,
\eeq
where we also have

\beq\label{6Pom55}
\prod_{i=1}^{6}\,P_{q-q}(q_i)\,\,\rightarrow\,\,
\prod_{i=1}^{6}\,P_{q-q}(b_i)\,\,.
\eeq
Putting in \eq{6Pom5} substitutions of \eq{6Pom2} we
obtain for the exponents of \eq{6Pom5}:

\begin{eqnarray}\label{6Pom6}
\,&\,&\,e^{i\,\vec{x}_{1}(-2\vec{b}+2\vec{b}_{2}/3+2\vec{b}_{3}/3
+\vec{b}_4-\vec{b}_{1}/3)}\,\,e^{i\,\vec{x}_{2}(
\vec{b}_4-\vec{b}_1)}\,
e^{i\,\vec{x}_{3}(\vec{b}_{1}/2-\vec{b}_{2}/2)}\,
e^{i\,\vec{x}_{4}(-\vec{b}_{2}/3+2\vec{b}_{3}/3
-\vec{b}_{1}/3)}\,\\
\,&\,&\,
e^{i\,\vec{x}_{5}(-\vec{b}_{2}-\vec{b}_{4}
+\vec{b}_{5}+\vec{b}_{1})}\,
e^{i\,\vec{x}_{6}(-\vec{b}_{3}-\vec{b}_{4}
+\vec{b}_{6}+\vec{b}_{1})}\,\,.
\end{eqnarray}
The vertices of  \eq{6Pom3} and \eq{6Pom4} have no dependence
on $\,x_5\,$ and $\,x_6\,$ variables. Therefore, in integration
over $\,x_5\,$ and $\,x_6\,$ we obtain the following
delta functions:

\beq\label{6Pom7}
\delta^2(b_5+b_1-b_2-b_4)\,\,\rightarrow\,\,
b_5=b_2+b_4-b_1\,\,,
\eeq
and 
\beq\label{6Pom8}
\delta^2(b_6+b_1-b_3-b_4)\,\,\rightarrow\,\,
b_6=b_3+b_4-b_1\,\,.
\eeq
In this case, after the integration over $\,b_5\,$
and $\,b_6\,$,
the answer  for the functions $\,P_{q-q}(b_i)\,$( see \eq{6Pom55}):
looks as follows 
\beq\label{6Pomp9}
\prod_{i=1}^{6}\,P_{q-q}(q_i)\,\,\rightarrow\,\,
\prod_{i=1}^{4}\,P_{q-q}(b_i)\,P_{q-q}(
b_2+b_4-b_1)\,P_{q-q}(b_3+b_4-b_1)\,,
\eeq
Of course, if we need,  we should also replace $b_5$
and $b_6$ in the other exponents of \eq{6Pom6} 
substituting \eq{6Pom7}-\eq{6Pom8}.In our particular diagram
this replacement has no place since exponents of \eq{6Pom6} do not
depend on $b_5$ and $b_6$.

 We are ready to write the answer for our diagram. 
Performing a simple integration over variables $\,x_i\,$, where $\,i=1-4\,\,$, 
we obtain (we use
Gaussian functions of \eq{6Pom3} and \eq{6Pom4} with the 
four exponents of \eq{6Pom6}):

\begin{eqnarray}\label{6Pom10}
D_{1,6Pom}(b)\,& = &\,
\frac{6\,\alpha^4}{\pi^4}\,
\int\,\prod_{i=1}^{4}\,d^2\,b_i\,P_{q-q}(b_i)\,
P_{q-q}(
b_2+b_4-b_1)\,P_{q-q}(b_3+b_4-b_1)\,\\
\,&\,&\,
e^{-\frac{3\,\alpha}{2}\,(2\vec{b}-2\vec{b}_{2}/3-
2\vec{b}_{3}/3-\vec{b}_{4}+\vec{b}_{1}/3 )^2\,-
\,\frac{\alpha}{2}\,(\vec{b}_{4}-\vec{b}_{1})^2\,-\,
\frac{\alpha}{2}\,(\vec{b}_{1}-\vec{b}_{2})^2\,-\,
\frac{2\,\alpha}{3}\,(\vec{b}_{3}-\vec{b}_{2}/2-
\vec{b}_{1}/2)^2\,} 
\end{eqnarray}

 All other diagrams, which contribute to the elastic amplitude,
we calculate using the same methods which have been described above. 
The answer for 
entire amplitude, which included expression for
all possible diagrams of our model is written in Appendix B.

\section*{Appendix B:}

 \renewcommand{\theequation}{B.\arabic{equation}}
\setcounter{equation}{0}

\begin{figure}[hptb]
\begin{center}
\psfig{file=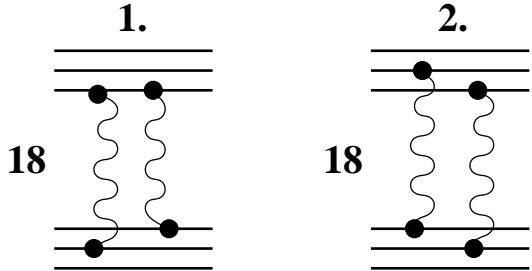,width=70mm}
\end{center} 
\caption{\it\large Two diagrams for the interactions of two quark pairs.}
\label{App7}
\end{figure}

 In this appendix we present the  resulting  expressions for the full
elastic amplitude, order by order. 
We have for the amplitude ( see \eq{Ampli1}):
\beq\label{B1}
A(s,b)\,=\,A_{1pair}(s,b)-A_{2pairs}(s,b)+A_{3pairs}(s,b)+\,...\,+\,A_{9pairs}(s,b).
\eeq
The answer for the contribution to the amplitude from the
one pair of quarks , $\,A_{1pair}(s,b)\,$,
is given in \eq{Ampli2}. So, we start from the 
two pairs contribution. The diagrams of this contribution 
are shown in  Fig.~\ref{App7}.

We have for $\,A(s,b)_{2pair}\,$:
\begin{eqnarray}\label{B2}
A_{2pairs}(s,b)\,&=&\,\frac{54\alpha^2}{5\pi^2}\,
\int\,\prod_{i=1}^{2}\,d^2\,b_i\,P_{q-q}(b_i)\,
e^{-\frac{6\alpha}{5}(\vec{b}_{1}/2+\vec{b}_{2}/2-\vec{b})^2\,-\,
\frac{\alpha}{2}(\vec{b}_{1}-\vec{b}_{2})^2}\,+\,\\
\,&\,+\,&\,
\,\frac{27\alpha^2}{2\pi^2}\,
\int\,\prod_{i=1}^{2}\,d^2\,b_i\,P_{q-q}(b_i)\,
e^{-3\alpha(\vec{b}_{1}/2+\vec{b}_{2}/2-\vec{b})^2\,-\,
\frac{\alpha}{4}(\vec{b}_{1}-\vec{b}_{2})^2}\,
\end{eqnarray} 

 The diagrams for the three pairs of interacting  quarks are shown
in  Fig.~\ref{App3}. We have for this contribution:
\begin{eqnarray}\label{B3}
A_{3pairs}(s,b)\,&=&\,\frac{6\alpha^3}{2\pi^3}
\int\prod_{i=1}^{3}d^2\,b_i\,P_{q-q}(b_i)
e^{-\frac{3\alpha}{2}(\vec{b}-\vec{b}_{1}/3-
\vec{b}_{2}/3-\vec{b}_{3}/3\,)^2 -\frac{\alpha}{2}
(\vec{b}_{2}-\vec{b}_{3})^2 -
\frac{2\alpha}{3}(\vec{b}_{1}-\vec{b}_{2}/2-
\vec{b}_{3}/2)^2}+\\
&\,+\,&\,\frac{27\,\alpha^3}{\pi^3}\,
\int\,\prod_{i=1}^{3}\,d^2\,b_i\,P_{q-q}(b_i)\,
e^{-\frac{\alpha}{2}\,(3\vec{b}-\vec{b}_{1}-
\vec{b}_{2}-\vec{b}_{3})^2\,-\,\frac{\alpha}{2}\,
(\vec{b}_{2}-\vec{b}_{1})^2\,-\,
\frac{3\,\alpha}{4}\,(\vec{b}-\vec{b}_{3})^2\,}\,+\,\\
&\,+\,&\,
\frac{9\,\alpha^3}{\pi^3}\,
\int\,\prod_{i=1}^{2}\,d^2\,b_i\,P_{q-q}(b_i)\,
\,P_{q-q}(3b-b_2-b_1)
e^{-\frac{\alpha}{3}\,(3\vec{b}-3\vec{b}_{1}/2-
3\vec{b}_{2}/2)^2\,-\,\frac{\alpha}{4}\,
(\vec{b}_{2}-\vec{b}_{1})^2\,}\,+\,\\
&\,+\,&\,
\frac{27\,\alpha^3}{\pi^3}\,
\int\,\prod_{i=1}^{3}\,d^2\,b_i\,P_{q-q}(b_i)\,
e^{-3\alpha\,(\vec{b}-\vec{b}_{2}/2-
\vec{b}_{3}/2)^2\,-\,\frac{\alpha}{2}\,
(\vec{b}_{2}-\vec{b}_{1})^2\,-\,
\frac{\alpha}{2}\,(\vec{b}_{1}-\vec{b}_{3})^2\,}\,
\end{eqnarray}

\begin{figure}[hptb]
\begin{center}
\psfig{file=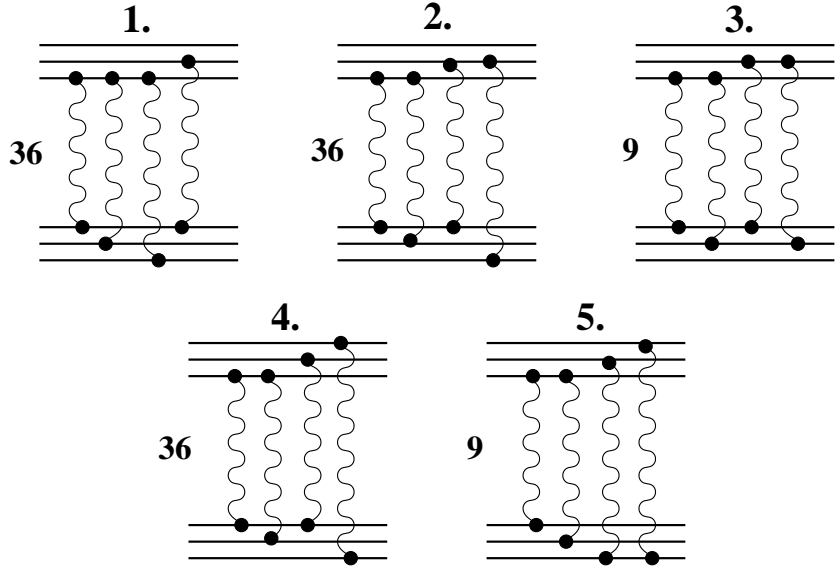,width=110mm}
\end{center} 
\caption{\it\large Five diagrams for interactions of four quark pairs.}
\label{App8}
\end{figure}

  For  four pairs amplitude we have the diagrams
of  Fig.~\ref{App8}.

For this contribution we have:

\begin{eqnarray}\label{B4}
A_{4pairs}(s,b)\,&=&\,\frac{36\alpha^4}{\pi^4}
\int\prod_{i=1}^{4}d^2\,b_i\,P_{q-q}(b_i)\,\\
&\,&
e^{-\frac{\alpha}{6}(6\vec{b}+\vec{b}_{1}-
2\vec{b}_{2}-2\vec{b}_{3}-3\vec{b}_{4})^2 
-\frac{\alpha}{2}(\vec{b}_{1}-\vec{b}_{4})^2 -
\,\frac{\alpha}{2}(\vec{b}_{1}-\vec{b}_{2})^2 -
\frac{3\alpha}{2}(\vec{b}_{1}/3+\vec{b}_{2}/3-
2\vec{b}_{3}/3)^2}+\\
&+&
\frac{36\alpha^4}{\pi^4}
\int\prod_{i=1}^{4}d^2\,b_i P_{q-q}(b_i)\,\\
&\,&
e^{-\frac{\alpha}{6}(6\vec{b}-\vec{b}_{1}-
2\vec{b}_{2}-\vec{b}_{3}-2\vec{b}_{4})^2 
-\frac{\alpha}{2}(\vec{b}_{1}-\vec{b}_{3})^2 -
\,\frac{\alpha}{2}(\vec{b}_{1}-\vec{b}_{2})^2 -
\frac{2\alpha}{3}(\vec{b}_{1}/2-\vec{b}_{2}/2-
\vec{b}_{3}+\vec{b}_{4})^2}+\\
&+&\,
\frac{27\alpha^4}{4\pi^4}
\int\prod_{i=1}^{3}\,d^2\,b_i\,P_{q-q}(b_i)\,
P_{q-q}(b_1-b_2-b_3)\,\\
&\,&
e^{-\frac{3\alpha}{4}(2\vec{b}-
\vec{b}_{2}-\vec{b}_{3})^2 
\,-\,\frac{\alpha}{2}(\vec{b}_{1}-\vec{b}_{3})^2\,-\,
\,\frac{\alpha}{2}(\vec{b}_{1}-\vec{b}_{2})^2}+\\
&\,+\,&\,
\frac{27\alpha^4}{4\pi^4}
\int\prod_{i=1}^{3}\,d^2\,b_i\,P_{q-q}(b_i)\,
P_{q-q}(3b-b_2-b_3)\,\\
&\,&
e^{-\frac{3\alpha}{4}(2\vec{b}-
\vec{b}_{2}-\vec{b}_{3})^2 
\,-\,\frac{\alpha}{2}(\vec{b}_{1}-\vec{b}_{3})^2\,-\,
\,\frac{\alpha}{2}(\vec{b}_{1}-\vec{b}_{2})^2}+\\
&\,+\,&\,
\frac{9\alpha^4}{\pi^4}
\int\prod_{i=1}^{4}\,d^2\,b_i\,P_{q-q}(b_i)\,\\
&\,&
e^{-\frac{3\alpha}{4}(4\vec{b}-\vec{b}_{1}-
\vec{b}_{2}-\vec{b}_{3}-\vec{b}_{4})^2 
\,-\,\frac{\alpha}{2}(\vec{b}_{4}-\vec{b}_{3})^2\,-\,
\,\frac{\alpha}{2}(\vec{b}_{1}-\vec{b}_{2})^2\,-\,
\frac{\alpha}{12}(\vec{b}_{1}+\vec{b}_{2}-
-\vec{b}_{3}-\vec{b}_{4})^2}\,.
\end{eqnarray}

\begin{figure}[hptb]
\begin{center}
\psfig{file=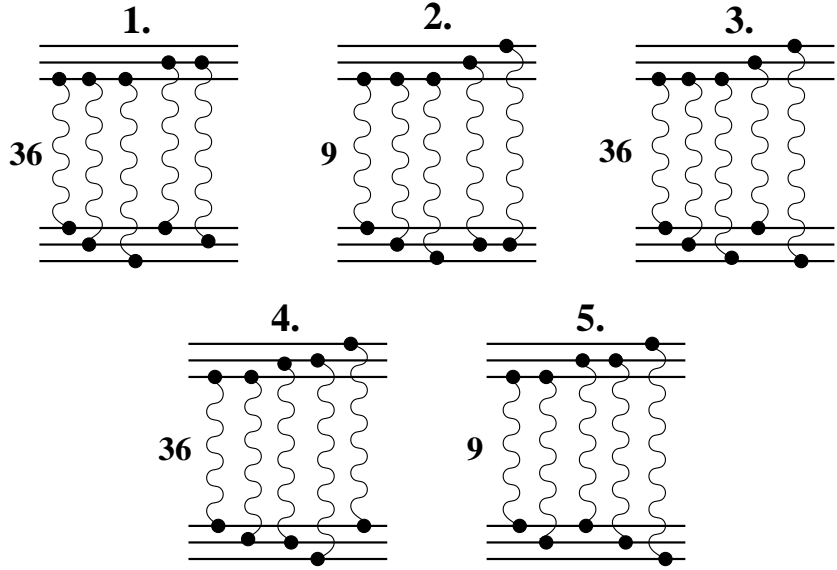,width=110mm}
\end{center} 
\caption{\it\large Five diagrams for the interactions of five quark pairs.}
\label{App9}
\end{figure}

  The five pairs interaction diagrams are shown in
 Fig.~\ref{App9}.

In this case we have:

\begin{eqnarray}\label{B5}
A_{5pairs}(s,b)\,&=&\,\frac{36\alpha^4}{\pi^4}
\int\prod_{i=1}^{4}d^2\,b_i\,P_{q-q}(b_i)
\,P_{q-q}(b_2+b_4-b_1)\,\\
&\,&
e^{-\frac{3\alpha}{2}(2\vec{b}+\vec{b}_{1}/3-
2\vec{b}_{2}/3-2\vec{b}_{3}/3-\vec{b}_{4})^2 
-\frac{\alpha}{2}(\vec{b}_{1}-\vec{b}_{4})^2 -
\,\frac{\alpha}{2}(\vec{b}_{1}-\vec{b}_{2})^2 -
\frac{2\alpha}{3}(\vec{b}_{1}+\vec{b}_{2}-\vec{b}_{3})^2}+\\
&+&
\frac{9\alpha^4}{\pi^4}
\int\prod_{i=1}^{4}d^2\,b_i\,P_{q-q}(b_i)
\,P_{q-q}(3b-b_1+b_2-b_3-b_4)\,\\
&\,&
e^{-\frac{2\alpha}{3}(3\vec{b}/2-\vec{b}_{1}/2+
\vec{b}_{2}-3\vec{b}_{3}/2-\vec{b}_{4}/2)^2 
-\frac{\alpha}{2}(\vec{b}_{2}-\vec{b}_{4})^2 -
\,\frac{\alpha}{2}(3\vec{b}-\vec{b}_{1}-\vec{b}_{3}-
\vec{b}_{4})^2 -
\frac{2\alpha}{3}(\vec{b}_{1}-\vec{b}_{2}/2-
\vec{b}_{4}/2)^2}+\\
&+&
\frac{36\alpha^4}{\pi^4}
\int\prod_{i=1}^{4}d^2\,b_i\,P_{q-q}(b_i)
\,P_{q-q}(3b-b_2-b_4)\,\\
&\,&
e^{-\frac{2\alpha}{3}(3\vec{b}/2-\vec{b}_{1}/2-
\vec{b}_{2}-\vec{b}_{3}/2-\vec{b}_{4})^2 
-\frac{\alpha}{2}(\vec{b}_{1}-\vec{b}_{4})^2 -
\,\frac{\alpha}{2}(\vec{b}_{1}-\vec{b}_{3})^2 -
\frac{2\alpha}{3}(\vec{b}_{1}/2+\vec{b}_{2}-\vec{b}_{3}-
\vec{b}_{4}/2)^2}+\\
&+&
\frac{36\alpha^4}{\pi^4}
\int\prod_{i=1}^{4}d^2\,b_i\,P_{q-q}(b_i)
\,P_{q-q}(3b-b_2-b_4)\,\\
&\,&
e^{-\frac{2\alpha}{3}(3\vec{b}-\vec{b}_{1}-
\vec{b}_{2}/2-\vec{b}_{3}/2-\vec{b}_{4})^2 
-\frac{\alpha}{2}(\vec{b}_{1}-\vec{b}_{2})^2 -
\,\frac{\alpha}{2}(\vec{b}_{2}-\vec{b}_{3})^2 -
\frac{2\alpha}{3}(\vec{b}_{1}/2-\vec{b}_{2}/2+\vec{b}_{3}-
\vec{b}_{4})^2}+\\
&+&
\frac{27\alpha^3}{16\,\pi^3}
\int\prod_{i=1}^{3}d^2\,b_i\,P_{q-q}(b_i)
\,P_{q-q}(3b-b_2-b_3)\,
\,P_{q-q}(b_1-b_2-b_3)\,\\
&\,&
e^{-\frac{3\alpha}{4}(2\vec{b}-
\vec{b}_{2}-\vec{b}_{3})^2 
-\frac{\alpha}{2}(\vec{b}_{1}-\vec{b}_{2})^2 -
\,\frac{\alpha}{4}(\vec{b}_{2}-\vec{b}_{3})^2 -
\frac{\alpha}{3}(\vec{b}_{1}-\vec{b}_{2}/2-\vec{b}_{3}/2)^2}\,.
\end{eqnarray}

 The diagrams with the interaction of six quark pairs are shown
in  Fig.~\ref{App6}.
For this contribution we have:
\begin{eqnarray}
A_{6pairs}(s,b)\,&=&\,\frac{6\,\alpha^4}{\pi^4}\,
\int\,\prod_{i=1}^{4}\,d^2\,b_i\,P_{q-q}(b_i)\,
P_{q-q}(
b_2+b_4-b_1)\,P_{q-q}(b_3+b_4-b_1)\,\\
&\,&\,
e^{-\frac{3\alpha}{2}(2\vec{b}-2\vec{b}_{2}/3-
2\vec{b}_{3}/3-\vec{b}_{4}+\vec{b}_{1}/3 )^2\,-
\,\frac{\alpha}{2}(\vec{b}_{4}-\vec{b}_{1})^2\,-\,
\frac{\alpha}{2}(\vec{b}_{1}-\vec{b}_{2})^2\,-\,
\frac{2\,\alpha}{3}(\vec{b}_{3}-\vec{b}_{2}/2-
\vec{b}_{1}/2)^2\,}+\\
&+&
\frac{36\,\alpha^4}{\pi^4}\,
\int\,\prod_{i=1}^{4}\,d^2\,b_i\,P_{q-q}(b_i)\,
P_{q-q}(3b-b_2-b_4)\,
P_{q-q}(b_2+b_4-b_1)\,\\
&\,&\,
e^{-\frac{2\alpha}{3}\,(3\vec{b}-\vec{b}_{1}/2-
\vec{b}_{2}+\vec{b}_{3}-3\vec{b}_{4}/2 )^2\,-
\,\frac{\alpha}{2}\,(\vec{b}_{2}-\vec{b}_{1})^2\,-\,
\frac{\alpha}{2}\,(\vec{b}_{1}-\vec{b}_{4})^2\,-\,
\frac{2\,\alpha}{3}\,(\vec{b}_{3}-\vec{b}_{1}-
\vec{b}_{2})^2\,}+\\
&+&
\frac{36\,\alpha^4}{\pi^4}\,
\int\,\prod_{i=1}^{4}\,d^2\,b_i\,P_{q-q}(b_i)\,
P_{q-q}(3b+b_1-b_2-b_3-b_4)\,
P_{q-q}(b_2+b_4-b_1)\,\\
&\,&\,
e^{-\frac{2\alpha}{3}\,(3\vec{b}+\vec{b}_{1}/2-
\vec{b}_{2}-\vec{b}_{3}-3\vec{b}_{4}/2 )^2\,-
\,\frac{\alpha}{2}\,(\vec{b}_{2}-\vec{b}_{1})^2\,-\,
\frac{\alpha}{2}\,(\vec{b}_{1}-\vec{b}_{4})^2\,-\,
\frac{2\,\alpha}{3}\,(\vec{b}_{3}-\vec{b}_{1}/2-
\vec{b}_{2}/2)^2\,}+\\
&+&
\frac{6\,\alpha^4}{\pi^4}\,
\int\,\prod_{i=1}^{4}\,d^2\,b_i\,P_{q-q}(b_i)\,
P_{q-q}(3b-b_2-b_4)\,
P_{q-q}(3b-b_1-b_3)\,\\
&\,&\,
e^{-\frac{2\alpha}{3}\,(3\vec{b}-\vec{b}_{1}-
\vec{b}_{2}/2-\vec{b}_{3}/2-\vec{b}_{4})^2\,-
\,\frac{\alpha}{2}\,(\vec{b}_{2}-\vec{b}_{1})^2\,-\,
\frac{\alpha}{2}\,(\vec{b}_{2}-\vec{b}_{3})^2\,-\,
\frac{2\,\alpha}{3}\,(\vec{b}_{4}-
\vec{b}_{3}-\vec{b}_{1}/2+\vec{b}_{2}/2)^2\,}\,.
\end{eqnarray}

\begin{figure}[hptb]
\begin{center}
\psfig{file=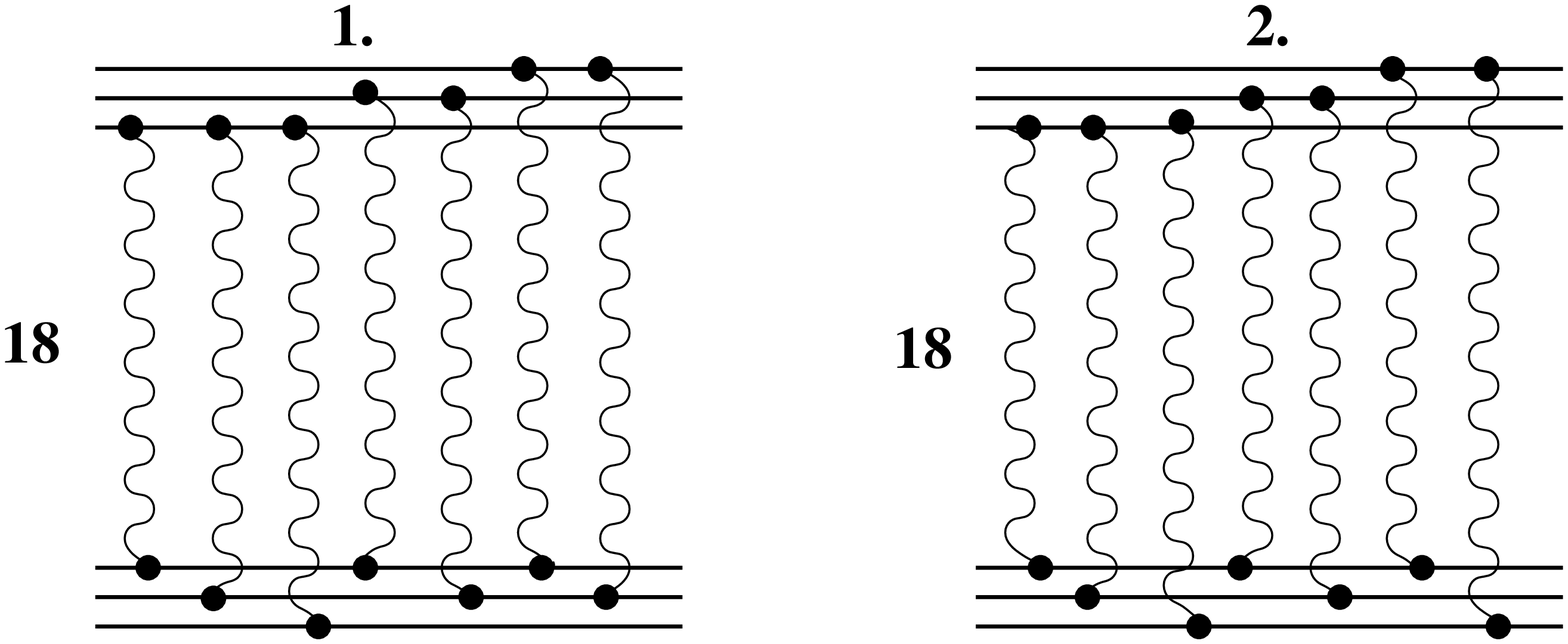,width=110mm}
\end{center} 
\caption{\it\large Two diagrams for the interactions of seven quark pairs.}
\label{App10}
\end{figure}

  The diagrams with the interaction of  seven quark pairs are shown in
 Fig.~\ref{App10}.

For these diagrams we have :

\begin{eqnarray}
A_{7pairs}(s,b)\,&=&\,\frac{18\,\alpha^4}{\pi^4}\,
\int\,\prod_{i=1}^{4}\,d^2\,b_i\,P_{q-q}(b_i)\,\\
&\,&
P_{q-q}(2b-b_4-b_3)\,
P_{q-q}(b_2+b_4-b_1)\,
P_{q-q}(b_1-b_2+b_3)\\
&\,&\,
e^{-\frac{2\alpha}{3}(3\vec{b}-\vec{b}_{1}/2-
\vec{b}_{2}/2-\vec{b}_{3}-\vec{b}_{4})^2\,-
\,\frac{\alpha}{2}(\vec{b}_{4}-\vec{b}_{1})^2\,-\,
\frac{\alpha}{2}(\vec{b}_{1}-\vec{b}_{2})^2\,-\,
\frac{2\,\alpha}{3}(\vec{b}_{3}-\vec{b}_{2}+
\vec{b}_{1}/2-\vec{b}_{4}/2)^2\,}+\\
&+&
\frac{18\,\alpha^4}{\pi^4}\,
\int\,\prod_{i=1}^{4}\,d^2\,b_i\,P_{q-q}(b_i)\,\\
&\,&
P_{q-q}(3b-b_4-b_3)\,
P_{q-q}(3b-b_4-b_2)\,
P_{q-q}(b_2-b_1+b_4)\\
&\,&\,
e^{-\frac{2\alpha}{3}(3\vec{b}+3\vec{b}_{1}/2-
\vec{b}_{2}-\vec{b}_{3}-3\vec{b}_{4}/2)^2\,-
\,\frac{\alpha}{2}(\vec{b}_{4}-\vec{b}_{1})^2\,-\,
\frac{\alpha}{2}(\vec{b}_{1}-\vec{b}_{2})^2\,-\,
\frac{2\,\alpha}{3}(\vec{b}_{3}-\vec{b}_{2}/2-
\vec{b}_{1}/2)^2\,}\,.
\end{eqnarray}

\begin{figure}[hptb]
\begin{center}
\psfig{file=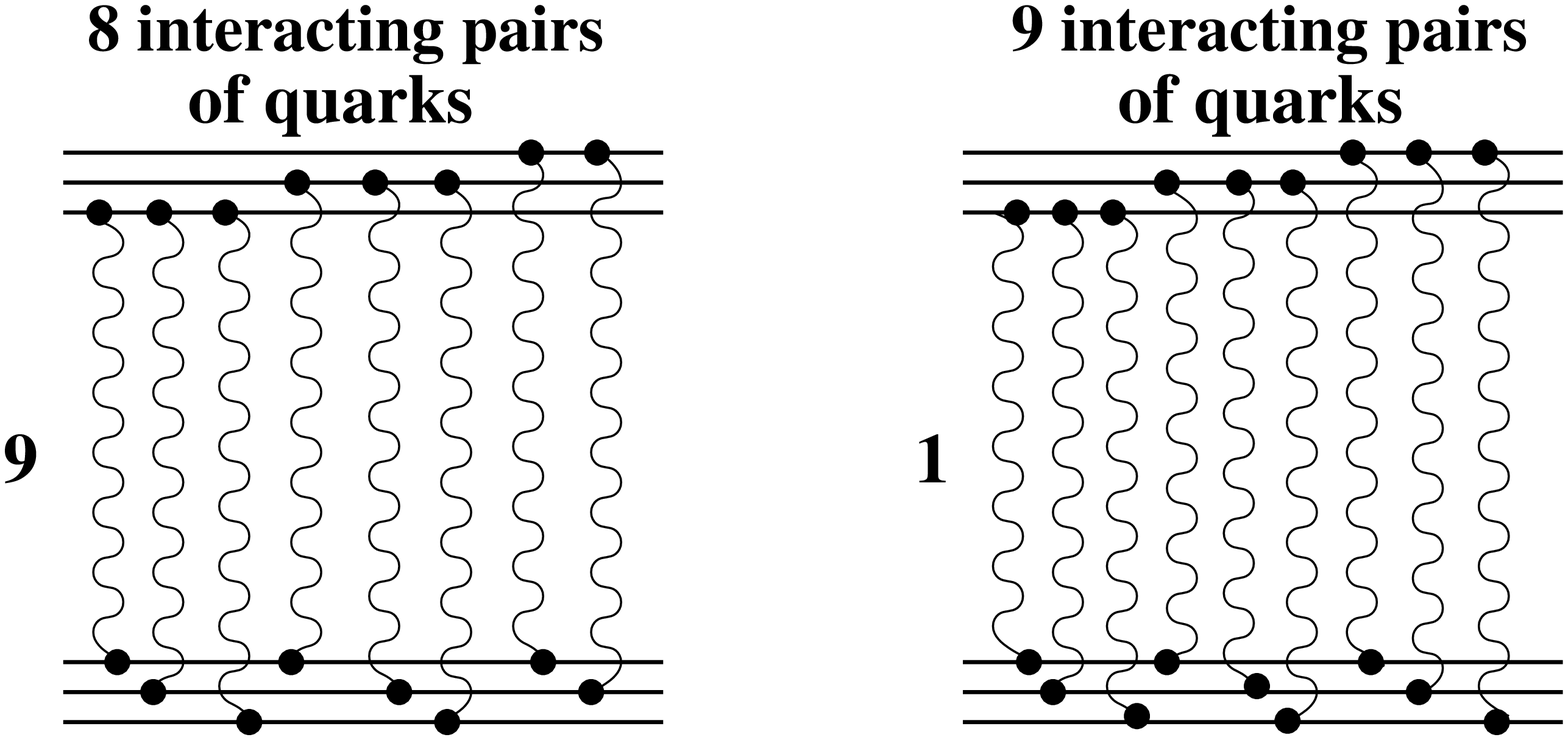,width=110mm}
\end{center} 
\caption{\it\large Two diagrams  for  the interactions of eight and nine quark pairs.}
\label{App11}
\end{figure}

  Finally, there are only one type of diagrams with the eight
and nine quark pairs interactions, see Fig.~\ref{App11}.

So we have for the diagrams with the interaction of  eight quark pairs:

\begin{eqnarray}
A_{8pairs}(s,b)\,&=&\,\frac{\alpha^4}{\pi^4}\,
\int\,\prod_{i=1}^{4}\,d^2\,b_i\,P_{q-q}(b_i)\,
P_{q-q}(2b-b_1)\,
P_{q-q}(2b-b_2)\,\\
&\,&
P_{q-q}(2b-b_1-b_2-b_3)\,
P_{q-q}(b_1-b_2+b_4)\\
&\,&\,
e^{-\frac{2\alpha}{3}(\vec{b}-\vec{b}_{1}/2+
\vec{b}_{2}/2-\vec{b}_{4})^2\,-
\,\frac{\alpha}{2}(\vec{b}_{2}-\vec{b}_{1})^2\,-\,
\frac{\alpha}{2}(2\vec{b}-
\vec{b}_{1}-\vec{b}_{2})^2\,-\,
\frac{2\,\alpha}{3}(2\vec{b}-\vec{b}_{1}/2-
\vec{b}_{2}/2-\vec{b}_{3})^2\,}\,,
\end{eqnarray}
and the diagram for  nine quark pair interactions gives: 

\begin{eqnarray}
A_{9pairs}(s,b)\,&=&\,\frac{\alpha^4}{9\,\pi^4}\,
\int\,\prod_{i=1}^{4}\,d^2\,b_i\,P_{q-q}(b_i)\,
P_{q-q}(2b-b_1)\,
P_{q-q}(2b-b_4)\,\\
&\,&
P_{q-q}(2b-b_1+b_2-b_4)
P_{q-q}(b_3-b_2+b_4)
P_{q-q}(b_1-b_2-b_3)\\
&\,&\,
e^{-\frac{2\alpha}{3}(\vec{b}+\vec{b}_{1}/2-
\vec{b}_{3}-\vec{b}_{4}/2)^2\,-
\,\frac{\alpha}{2}(\vec{b}_{4}-\vec{b}_{1})^2\,-\,
\frac{\alpha}{2}(2\vec{b}-
\vec{b}_{1}-\vec{b}_{4})^2\,-\,
\frac{2\,\alpha}{3}(\vec{b}_{1}/2-
\vec{b}_{2}-\vec{b}_{4}/2)^2\,}\,.
\end{eqnarray}

 In our numerical calculations we used only the
$A(s,b)_{1pair}$-$A(s,b)_{7pairs}\,$ terms of the amplitude.
The term
$A(s,b)_{6pairs}$-$A(s,b)_{7pairs}\,$ works
 only at energies close to the LHC one.

\section*{Appendix C:}

\renewcommand{\theequation}{C.\arabic{equation}}
\setcounter{equation}{0}

  As example of expression for the calculation of the survival probability (SP)
in the framework of our approach we consider the term of elastic amplitude
with three interacting quark pairs:

\begin{eqnarray}\label{C1}
A(s,b)_{3pairs}\,&=&\,\frac{6\alpha^3}{2\pi^3}
\int\prod_{i=1}^{3}d^2\,b_i\,P_{q-q}(b_i)
e^{-\frac{3\alpha}{2}(\vec{b}-\vec{b}_{1}/3-
\vec{b}_{2}/3-\vec{b}_{3}/3\,)^2 -\frac{\alpha}{2}
(\vec{b}_{2}-\vec{b}_{3})^2 -
\frac{2\alpha}{3}(\vec{b}_{1}-\vec{b}_{2}/2-
\vec{b}_{3}/2)^2}+\\
&\,+\,&\,\frac{27\,\alpha^3}{\pi^3}\,
\int\,\prod_{i=1}^{3}\,d^2\,b_i\,P_{q-q}(b_i)\,
e^{-\frac{\alpha}{2}\,(3\vec{b}-\vec{b}_{1}-
\vec{b}_{2}-\vec{b}_{3})^2\,-\,\frac{\alpha}{2}\,
(\vec{b}_{2}-\vec{b}_{1})^2\,-\,
\frac{3\,\alpha}{4}\,(\vec{b}-\vec{b}_{3})^2\,}\,+\,\\
&\,+\,&
\frac{9\alpha^3}{\pi^3}
\int\,\prod_{i=1}^{2}\,d^2\,b_i\,P_{q-q}(b_i)
P_{q-q}(3b-b_2-b_1)
e^{-\frac{\alpha}{3}(3\vec{b}-3\vec{b}_{1}/2-
3\vec{b}_{2}/2)^2 -\frac{\alpha}{4}
(\vec{b}_{2}-\vec{b}_{1})^2}+\\
&\,+\,&\,
\frac{27\,\alpha^3}{\pi^3}\,
\int\,\prod_{i=1}^{3}\,d^2\,b_i\,P_{q-q}(b_i)\,
e^{-3\alpha\,(\vec{b}-\vec{b}_{2}/2-
\vec{b}_{3}/2)^2\,-\,\frac{\alpha}{2}\,
(\vec{b}_{2}-\vec{b}_{1})^2\,-\,
\frac{\alpha}{2}\,(\vec{b}_{1}-\vec{b}_{3})^2\,}\,
\end{eqnarray}

Let us rewrite the first term of this expression. Using receipt of the
\eq{Sp10}, we  rewrite:

\begin{eqnarray}\label{C2}
\prod_{i=1}^{3}\,P_{q-q}(b_i)\,&\,\rightarrow\,&\,
F_{SP,3pairs}^{1}(b_1,b_2,b_3)\,=\,
\frac{e^{-\frac{b^{2}}{2\,R^{2}_{Q-H}}}}
{2\pi\,R^{2}_{H}}\hat{P}_{q-q}(b_1)\,P_{q-q}(b_2)\,
P_{q-q}(b_3)\,+\,\\
& + &
\frac{e^{-\frac{b^{2}}{2\,R^{2}_{Q-H}}}}
{2\pi\,R^{2}_{H}}\hat{P}_{q-q}(b_2)P_{q-q}(b_3)
P_{q-q}(b_1)+
\frac{e^{-\frac{b^{2}}{2\,R^{2}_{Q-H}}}}
{2\pi\,R^{2}_{H}}\hat{P}_{q-q}(b_3)P_{q-q}(b_1)
P_{q-q}(b_2).
\end{eqnarray}
The expression of the r.h.s. with \eq{C2}
instead of $\prod_{i=1}^{3}\,P_{q-q}(b_i)\,$ in 
the first term of elastic amplitude.  \eq{C1}
gives the answer for the first term of SP amplitude
$\hat{A}_{3pairs}(s,b)$.
Of course, obtaining $\hat{A}(s,b)_{3pairs}$
we must to perform such replacement  in each term of
\eq{C1}. Doing so we obtain:

\beq\label{C3}
\hat{A}(s,b)_{3pairs}\,=\,
\eeq
$$
\,=\,\frac{6\alpha^3}{2\pi^3}
\int\prod_{i=1}^{3}d^2\,b_i\,F_{SP,3pairs}^{1}(b_1,b_2,b_3)\,
e^{-\frac{3\alpha}{2}(\vec{b}-\vec{b}_{1}/3-
\vec{b}_{2}/3-\vec{b}_{3}/3\,)^2 -\frac{\alpha}{2}
(\vec{b}_{2}-\vec{b}_{3})^2 -
\frac{2\alpha}{3}(\vec{b}_{1}-\vec{b}_{2}/2-
\vec{b}_{3}/2)^2}\,+
$$
$$
+\,\frac{27\,\alpha^3}{\pi^3}\,
\int\,\prod_{i=1}^{3}\,d^2\,b_i\,F_{SP,3pairs}^{2}(b_1,b_2,b_3)\,
e^{-\frac{\alpha}{2}\,(3\vec{b}-\vec{b}_{1}-
\vec{b}_{2}-\vec{b}_{3})^2\,-\,\frac{\alpha}{2}\,
(\vec{b}_{2}-\vec{b}_{1})^2\,-\,
\frac{3\,\alpha}{4}\,(\vec{b}-\vec{b}_{3})^2\,}\,+
$$
$$
+\,
\frac{9\,\alpha^3}{\pi^3}\,
\int\,\prod_{i=1}^{2}\,d^2\,b_i\,F_{SP,3pairs}^{3}(b_1,b_2)\,
\,P_{q-q}(3b-b_2-b_1)
e^{-\frac{\alpha}{3}\,(3\vec{b}-3\vec{b}_{1}/2-
3\vec{b}_{2}/2)^2\,-\,\frac{\alpha}{4}\,
(\vec{b}_{2}-\vec{b}_{1})^2\,}\,+
$$
$$
+\,
\frac{27\,\alpha^3}{\pi^3}\,
\int\,\prod_{i=1}^{3}\,d^2\,b_i\,F_{SP,3pairs}^{4}(b_1,b_2,b_3)\,
e^{-3\alpha\,(\vec{b}-\vec{b}_{2}/2-
\vec{b}_{3}/2)^2\,-\,\frac{\alpha}{2}\,
(\vec{b}_{2}-\vec{b}_{1})^2\,-\,
\frac{\alpha}{2}\,(\vec{b}_{1}-\vec{b}_{3})^2\,}\,.
$$

\end{document}